\documentclass[twocolumn,showpacs,preprintnumbers,amsmath,amssymb]{revtex4}

\usepackage{graphicx}
\usepackage{dcolumn}
\usepackage{bm}

\begin{document}

\preprint{APS/123-QED}

\title{Force transmission in a packing of pentagonal particles}
\author{Emilien Az\'ema, Farhang Radja\"\i, Robert Peyroux}
\affiliation{LMGC, CNRS - Universit\'e Montpellier II, Place Eug\`ene Bataillon, 
34095 Montpellier cedex 05, France.}
\email{azema@lmgc.univ-montp2.fr}

\author{Gilles Saussine}
\affiliation{Innovation and Research Departement of SNCF, 45 rue de Londres, 
75379 PARIS Cedex 08}

 \date{\today}

\begin{abstract}
We perform a detailed analysis of the contact force network in a dense confined packing 
of pentagonal particles simulated by means of the contact dynamics method.  
The effect of particle shape is evidenced by comparing the 
data from pentagon packing and from a packing with identical characteristics except for 
the circular shape of the particles. A counterintuitive 
finding of this work is that, under steady shearing, the pentagon packing develops a
lower structural anisotropy than the disk packing. We show that this weakness  is  
compensated by a higher force anisotropy, leading to enhanced shear strength of 
the pentagon packing.  We revisit ``strong" and ``weak" force networks 
in the pentagon packing,    
but our simulation data provide also evidence for  a large 
class of ``very weak" forces carried mainly by vertex-to-edge contacts.  
The strong force chains are mostly composed of edge-to-edge contacts with a  
marked zig-zag aspect and a decreasing exponential probability distribution as 
in a disk packing.    
\end{abstract}

\maketitle

\section{Introduction}
\label{intro}

Among singular features of granular media, force transmission 
has received particular interest during the last decade.  
The contact forces  in model granular media, as observed by experiments and numerical 
simulations, are highly inhomogeneous and their probability density functions (pdf's) are wide
 \cite{Liu1995a,Radjai1996,Coppersmith1996,Mueth1998a,Lovol1999, Bardenhagen2000, Antony2001,
Majmudar2005,Silbert2002}. 
The granular texture is generically anisotropic in two respects: 
1) The contact normal directions are not random; 2) The force average as 
a function of  contact normal direction is not uniform. 
The corresponding fabric and force anisotropies in shear are responsible 
for mechanical strength at the scale of the packing \cite{Radjai1998,Kruyt1996,Bathurst1988,Rothenburg1989}.  
Another interesting aspect, first analyzed  in Ref.  \cite{Radjai1998}   
is the fact that the forces organize themselves in two distinct classes which 
contribute differently to fabric anisotropy, shear stress, and dissipation. 
In particular, the shear stress is fully transmitted via a ``strong'' contact network, materialized by 
force ``chains". The stability is ensured by the antagonist role 
of  ``weak'' contacts which prop strong force chains\cite{Radjai1998,Staron2005}. 

The force transmission properties have been for the most part investigated  
 in the case of granular media composed of 
isometric (circular or spheric) particles. 
However, in various fields of science and engineering, 
the grains are seldom so "perfect". For example, elongated and platy 
shapes are encountered 
in biomaterials or pharmaceutical applications. Such shapes have unequal dimensions  
and induce thus a degree of anisotropy in the bulk behavior in addition 
to fabric and force anisotropies \cite{Ouadfel2001,Antony2004,Cambou2004,Nouguier-Lehon2003}. 
On the other hand, granular geomaterials are often composed of angular particles 
with plane faces as polyhedra. While rounded particles enhance flowability,  
angular shape is susceptible 
to enhance the shear strength, a factor of vital importance to  
civil-engineering applications \cite{Nouguier-Lehon2003,Alonso-Marroquin2002,Pena2006}. 
The railway ballast is a well-known case where particle shape must be optimized to 
avoid excessive differential settlement  under  vertical 
loading \cite{Saussine2006,Markland1981,Wu2000}. In such circumstances, 
the analysis of force transmission is a key to improve performance.       

In dealing with effects of particle shape, the issue   
is that a general quantitative description of 
particle morphology requires various shape parameters. 
For regular polygons in 2D, for instance, the only shape 
parameter is the number of sides (besides the diameter) 
whereas for irregular polygons more information is needed about  
the positions of the vertices in a reference system attached to the particle. 
In soil mechanics, angularity and roundedness are among basic parameters 
used to describe particle shapes \cite{Mitchell2005}.
As far as force transmission is concerned, at least two parameters seem to 
be most relevant: 1) shape anisotropy (anisometry), which contributes to the anisotropy of 
stress transmission \cite{Ouadfel2001}; 2) {\it facettedness}, 
which allows for extended (face to face, edge to face and edge to edge) contacts between particles 
leading possibly to  the formation of columnar structures within a granular assembly. 

In this paper, we consider one of the simplest possible shapes, namely 
regular pentagons. Among regular polygons, the 
pentagon has the lowest number of 
sides, corresponding to the least roundedness in this category, without the 
pathological space-filling properties of triangles and squares. 
We seek to isolate the effect of edge-to-edge contacts on force transmission 
by comparing the data with a packing of circular particles that, apart from the particle shape, is identical in all respects (preparation, friction coefficients, particle size distribution) to the pentagon packing. Both packings are subjected to biaxial compression simulated by means 
of the contact dynamics method.  The presence  of edge-to-edge contacts 
affects both quantitatively and qualitatively the microstructure and 
the overall behavior during shear. These contacts do not transmit torques, 
but they are able to accommodate force lines that are usually 
unsustainable in packings of disks.    

This paper is organized as follows. We first present in Section \ref{procedure} 
the numerical procedures and a 
brief technical introduction to the detection and treatment of edge-to-edge contacts 
in the framework of the contact dynamics method. In Section \ref{strength}, 
we compare stress-strain and volume-change characteristics. Then,  
In Sections \ref{texture} and \ref{force}, we analyze the texture and 
force transmission features. In Section \ref{sd}, we focus on the pentagon packing and 
we analyze the  
structure of force networks with vertex-to-edge and 
edge-to-edge contacts. The main results are summarized and discussed 
 in Section \ref{conclusion}.

\section{Numerical procedures}
\label{procedure}

The simulations were carried out by means of the contact dynamics (CD) 
method \cite{Jean1992,Moreau2004}. 
The CD method is based on implicit time integration of the equations 
of motion and a nonsmooth formulation of  
mutual exclusion and dry friction between particles.  
This method requires no elastic repulsive potential and no smoothing 
of the Coulomb friction law 
for the determination of forces. For this reason, the simulations can be 
performed with large time steps compared to molecular dynamics simulations.  
We used the platform LMGC90 which is a multipurpose software  
developed in Montpellier, capable of modeling a collection of 
deformable or undeformable particles of
various shapes \cite{Dubois2003}.

\subsection {Contact dynamics for polygons}
The particles are rigid polygons exerting normal and shear forces, $f_n$ and $f_t$,
respectively, on each other. We attribute a positive sign to compressive normal forces.
The relative normal velocity $u_n$ between two particles in contact is counted 
positive when they move away from each other.
Then, the condition of geometrical contact  between two  particles is expressed by  
the following mutually exclusive alternatives:
\begin{equation}
\begin{array}{ll}
f_n \geqslant 0 & u_n = 0 \\
f_n = 0               & u_n > 0.
\end{array}
\label{eq:1}
\end{equation}

In the same way, the Coulomb friction law involves three mutually exclusive conditions:
\begin{equation}
\begin{array}{cc}
f_t = -\mu f_n                                         &       u_t > 0 \\
-\mu f_n \leqslant f_t \leqslant \mu f_n &       u_t = 0 \\
f_t = \mu f_n                                          &       u_t < 0 \\
\end{array}
\label{eq:2}
\end{equation}
where $u_t$ is the sliding velocity at the contact and $\mu$ is the friction coefficient.
The unknown variables are particle velocities and contact forces.
These are calculated at each time step by taking into account the conservation of 
momenta,
the constraints expressed by (\ref{eq:1}) and (\ref{eq:2}), 
and the dissipation of kinetic energy during inelastic  collisions between particles (ref).
We use an iterative research algorithm based on a  nonlinear Gauss-Seidel scheme. 
The uniqueness is not guaranteed for perfectly rigid particles in absolute terms. However, by 
initializing each step of calculation with the forces calculated  in the 
preceding step, the set of admissible solutions shrinks to fluctuations which are basically 
below the numerical solution. Let us note that in molecular dynamics simulations, this 
``force history" is  encoded  by construction in the particle positions.      

The research algorithm is applied to a set of potential contacts, identified or 
updated in each step. 
The contact detection between two bodies  consists in looking 
for the overlaps of the portions of space they occupy. 
The treatment of the mechanical interaction requires additionally the identification of a common tangent plane 
(a line in 2D). 
Of course, contact may take place through a larger contact zone than a single point. 
Several algorithms exist for overlap determination    
between convex polygons   
\cite{Dubois2003,Saussine2006}.
In 2D simulations of the present paper, the detection of contact between 
two convex polygonal bodies was implemented through the so-called "shadow overlap method" 
devised by Moreau \cite{Dubois2003,Saussine2006}, with reliability and robustness tested in several 
years of previous applications to various states of granular materials 
\cite{Nouguier-Lehon2003,Cholet2003,Azema2006}. 

\begin{figure}
\includegraphics[width=8cm]{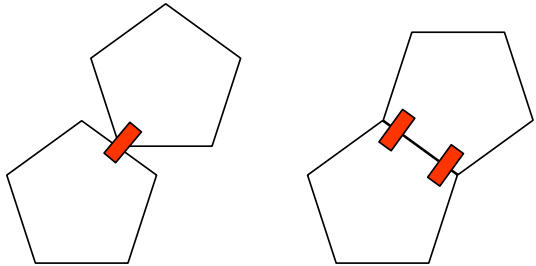}
\caption{Representation of simple (vertex-to-edge) and double 
(edge-to-edge) contacts between two pentagons.\label{fig01}}
\end{figure}

In  detection of contacts between two polygons,  two situations arise:  
1) If a single corner is found crossing an edge of the partner polygon, 
the direction of this edge is viewed as the tangent direction. 
By orthogonally projecting the intruding vertex onto the edge, 
one determines the penetration depth, while the nominal contact point 
is chosen at the center of this distance. 
Below, we will refer to this vertex-to-edge contact as ``simple" contact. 
2) In case of double intrusion, the common tangent line is fixed from as a mean  
between the two overlapping edges and  
a segment of this line is identified as the contact segment. 
The impenetrability between two particles at such an edge-to-edge 
contact is ensured by applying 
the contact laws (\ref{eq:1}) and  (\ref{eq:2}) to only two points of the contact segment (Fig. \ref{fig01}). 
For this reason, we refer below to edge-to-edge contacts as ``double" contacts.  
In practice, two forces are calculated at each double contact, but only their 
resultant and application point are material. In this respect, the choice of 
the two points representing a double contact does not affect the dynamics 
of the system.

\subsection{Numerical samples}

\begin{figure}
\includegraphics[width=8cm]{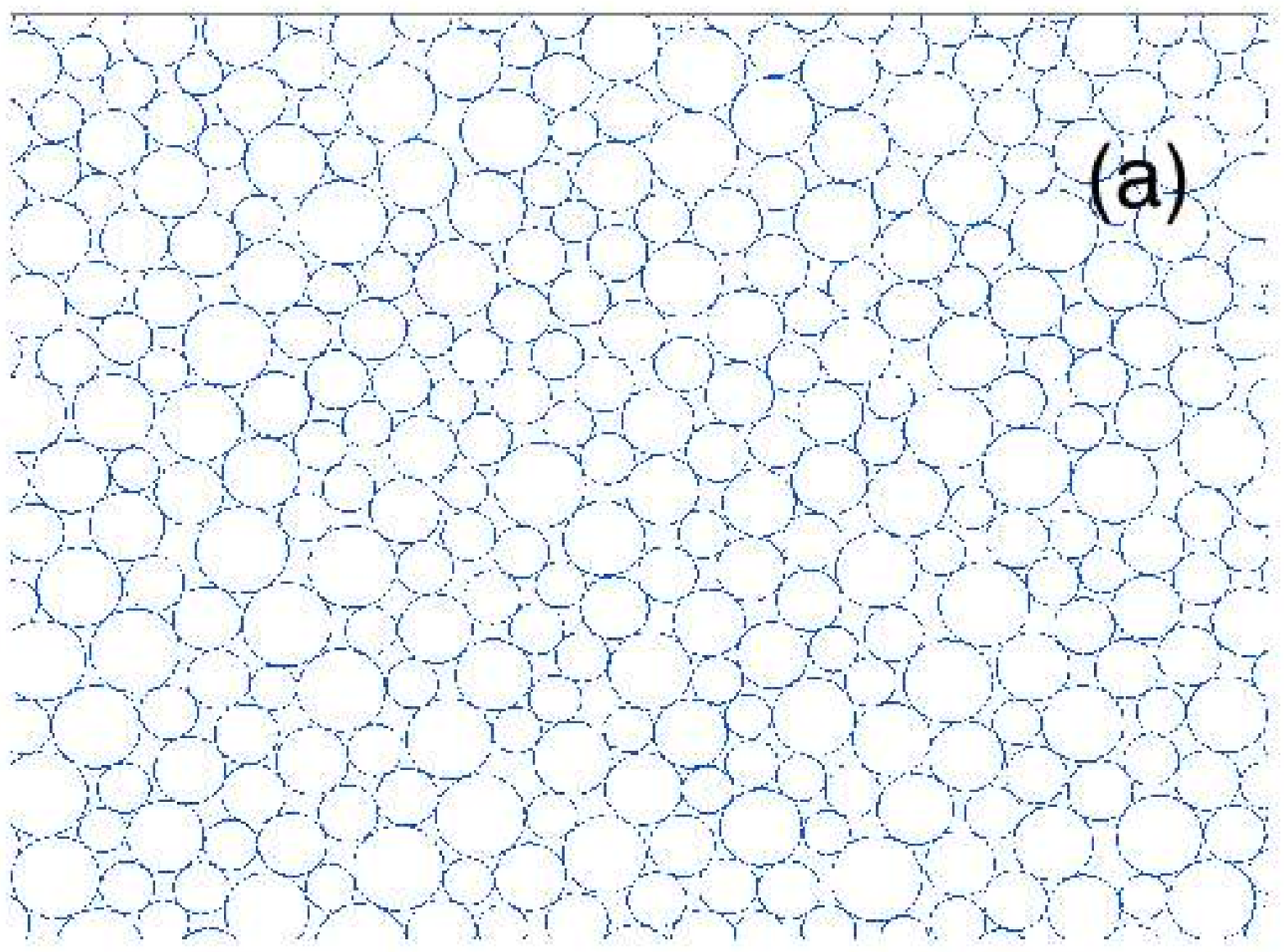}
\includegraphics[width=8cm]{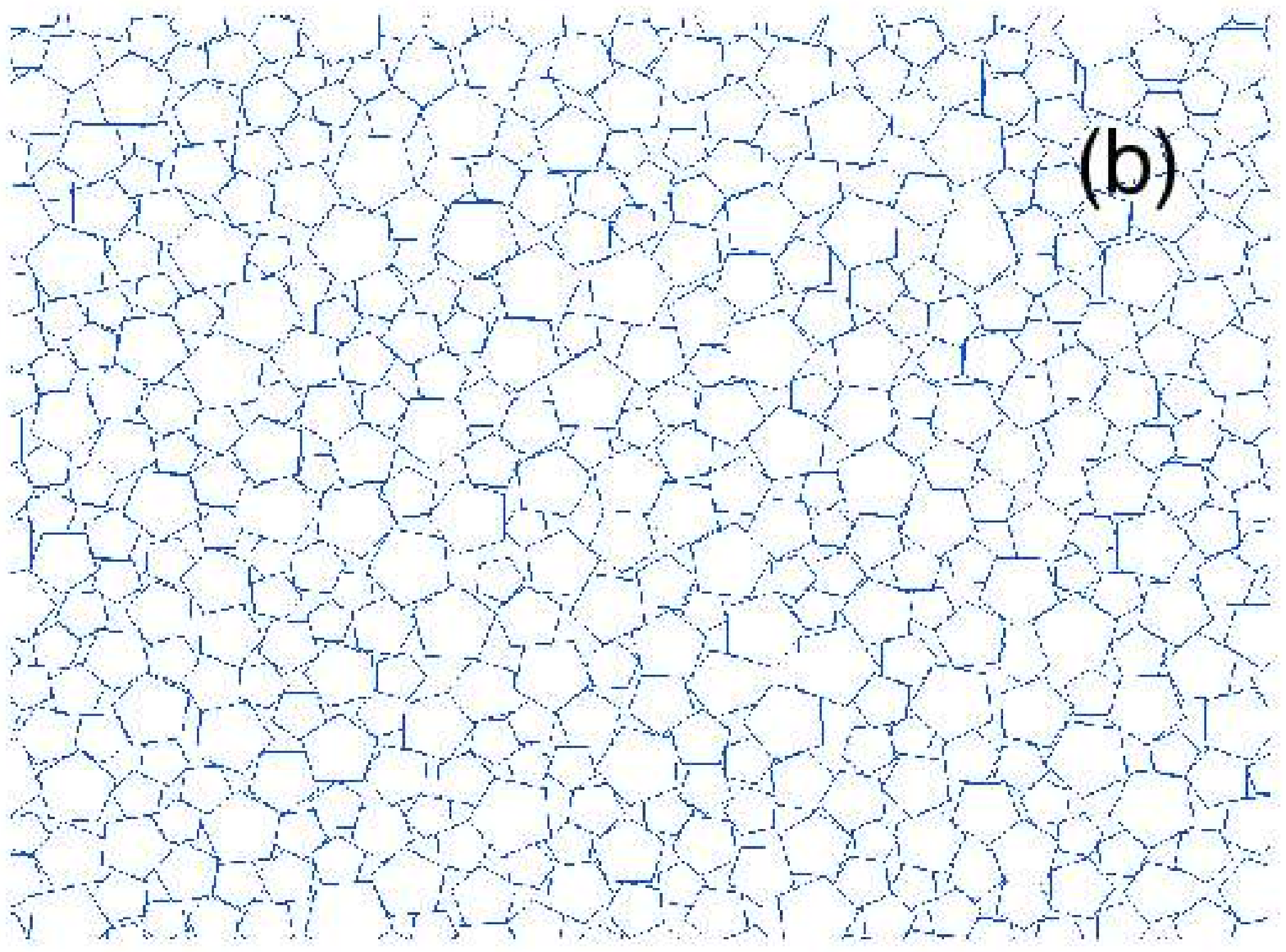}
\caption{Snapshots of a portion of the samples S2 (a) and S1 (b) composed of 
 circular and pentagonal particles, respectively.   \label{fig02}}
\end{figure}

We generated two  numerical samples. The first sample, denoted  
S1, is composed of 14400 regular pentagons of three different diameters: 
$50\%$ of diameter $2.5$ cm, $34\%$ of diameter $3.75$ cm and $16\%$ of diameter $5$ cm.
The second sample, denoted S2, is composed of 10000 discs 
with the same polydispersity. 
Both samples were prepared according to the same protocol. A dense 
packing was first constructed following simple geometrical rules \cite{Taboada2005}  
and then   compressed isotropically under a constant stress 
$\sigma_0=10^4$ Pa applied onto the right and top walls. 
The gravity was set to zero in order to 
avoid force gradients in the samples.
The coefficient of friction was set to 0.4 between grains and to 0 with the walls. 
At equilibrium, both numerical samples were in  isotropic stress state.  
The solid fraction was $\phi_0=0.80$ for S1 and  $\phi_0=0.82$ for S2. 
The aspect ratio was $h/l\approx2$, where $h$ and $l$ are the height and 
width of the sample, respectively. 
Figure \ref{fig02}  displays snapshots of the two packings  
at the end of isotropic compaction. 

The isotropic samples were subjected to vertical compression by downward 
displacement of the top wall at a constant velocity of $1$ cm/s for 
a constant confining stress $\sigma_0$ acting on the lateral walls.
The simulations were run up to a total cumulative vertical strain of $0.2$ with a time step 
of   $5.10^{-4}$ s.  The CPU time was $7.10^{-4}$ s and $5.10^{-4}$ s
 per particle and per time step on a G5 Apple computer.  
Since we are interested in quasistatic behavior, the shear rate should be 
such that the kinetic energy supplied by shearing is negligible compared to 
the static pressure. 
This can be formulated in terms of an "inertia parameter" $I$ \cite{GDR-MiDi2004} defined by 
\begin{equation}
I=\dot \varepsilon \sqrt{\frac{m}{p}},
\label{eq3}
\end{equation}
where $\dot \varepsilon=\dot y /y$ is the strain rate, 
$m$ is the total mass, and $p$ is the average pressure. The quasistatic limit is
characterized by the condition $I\ll1$. In our biaxial simulations, $I$ was 
below $10^{-3}$.  

\section{Strength and dilatancy}
\label{strength}

In this section, we compare the stress-strain and volume-change behavior 
between the packings of polygons (sample S1) and disks (sample S2). 
For the calculation of the stress tensor, we consider the "tensorial 
moment" ${\bm M}^i$ of each particle i defined by \cite{Moreau1997,Staron2005}:
\begin{equation}
M^i_{\alpha \beta} = \sum_{c \in i} f_{\alpha}^c r_{\beta}^c,
\label{eq:M}
\end{equation}
where  $f_{\alpha}^c$ is the $\alpha$ component of the force exerted on 
particle i at the contact c, $r_{\beta}^c$ is the $\beta$ component 
of the position vector of the same contact c, and the summation 
is runs over all contacts c of neighboring particles with the particle i (noted briefly by $c \in i$).
It can be shown that the tensorial moment of a collection of rigid particles is the sum of the 
tensorial moments of individual particles. 
The stress tensor ${\bm \sigma}$ for a packing of volume $V$  
is simply given by \cite{Moreau1997,Staron2005}: 
\begin{equation}
{\bm \sigma } = \frac{1}{V} \sum_{i \in V} {\bm M}^i =   \frac{1}{V}  \sum_{c \in V} f_{\alpha}^c \ell_{\beta}^c, 
\label{eq:M}
\end{equation}
where ${\bm \ell}^c$ is the intercenter vector joining the centers of the two touching particles at the 
contact $c$. Remark that the first summation runs over all particles whereas the second summation 
involves all contacts in the volume $V$, with each contact appearing only once.  
We extract the mean stress $p=(\sigma_1+\sigma_2)/2$, and the stress deviator $q=(\sigma_1-\sigma_2)/2$,
where $\sigma_1$ and $\sigma_2$ are the principal stresses. 
The major principal direction during vertical compression is vertical.

The strain parameters are the cumulative vertical, horizontal and 
volumetric strains $\varepsilon_1$, $\varepsilon_2$ and $\varepsilon_p$, respectively.
By definition, we have 
\begin{equation}
\varepsilon_1=\int \frac{dh}{h} = \ln \left( 1+ \frac{\Delta h}{h_0} \right),
\end{equation}
where $h_0$ is the initial height and $\Delta h = h_0 - h$ is the total downward displacement, and 
\begin{equation}
\varepsilon_p=\int \frac{dV}{V} = \ln \left( 1+ \frac{\Delta V}{V_0} \right),
\end{equation}
where $V_0$ is the initial volume and $\Delta V = V - V_0$ is the cumulative volume change.

\begin{figure}
\includegraphics[width=8cm]{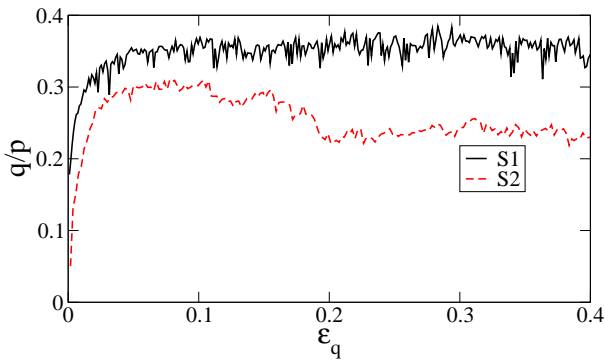}
\caption{Normalized shear stress $q/p$ as a function of cumulative shear strain $\varepsilon_q$ 
for the samples S1 and S2. \label{fig03}}
\end{figure}

Figure \ref{fig03} shows the normalized shear stress $q/p$ for the samples S1 and S2 
as a function of shear  strain  $\varepsilon_q \equiv \varepsilon_1 - \varepsilon_2$. 
For S2, we observe a classical behavior characterized by a hardening behavior 
followed by (slight) softening and a stress plateau corresponding to the 
residual state of soil mechanics \cite{Mitchell2005}. For S1, we observe no marked stress peak.  
The residual stress is higher for polygons ($\simeq 0.35$) than for disks ($ \simeq 0.28$). 
This means that the polygon packing has a higher angle of 
internal friction $\varphi$ defined by 
\begin{equation}
\sin \varphi = \frac{q}{p}. 
\end{equation}

\begin{figure}
\includegraphics[width=8cm]{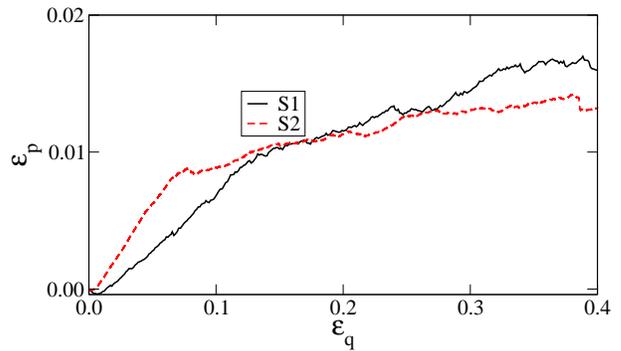}
\caption{Cumulative volumetric strain  
$\varepsilon_p$ as a function of cumulative shear strain $\varepsilon_q$ 
for the samples S1 and S2.  \label{fig04}}
\end{figure}

Figure \ref{fig04} displays the cumulative volumetric strain  
$\varepsilon_p$ for polygons and disks as a function of $\varepsilon_q$. 
Both samples dilate and tend to isochoric deformation at large strains.  
It is remarkable that the polygon packing S1 initially 
dilates less than the disk packing S2. This behavior is reversed at larger strains with a crossover 
occurring after the peak state. Notice that the solid fraction is initially lower in S1 ($0.80$) 
than in S2 ($0.82$). This is because it is more difficult to obtain a compact packing 
with polygonal shapes by isotropic compression as a result of enhanced steric effects compared to 
disks. In other words, angular particles can form larger pores compared to rounded particles.  
The volumetric deformation can also be expressed in terms 
of the so-called "dilatancy angle" $\psi$ defined by \cite{Wood1990}
\begin{equation}
\sin \psi = \frac{\varepsilon_p}{\varepsilon_q}. 
\end{equation}
The cumulative angle of dilatancy, i.e. during shear up to the residual state, 
is only slightly higher for the 
polygon packing than the disk packing. 

\begin{figure}
\includegraphics[width=8cm]{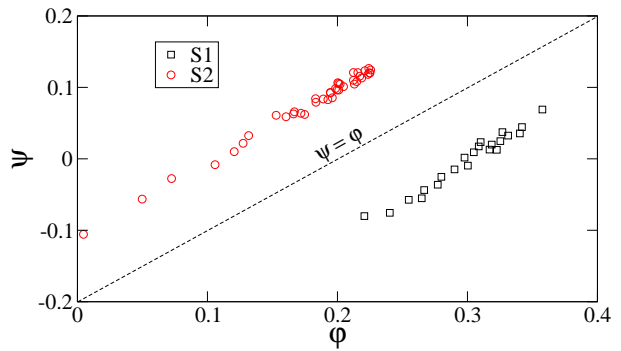}
\caption{Stress-dilatancy relation between 
dilatancy angle $\psi$ and internal angle of friction $\varphi$ for the 
samples S1 and S2. \label{fig05}}
\end{figure}

The plot of $\psi$ as a function of $\varphi$, i.e. the so-called  stress-dilatancy diagram,  
is shown in Fig. \ref{fig05} for polygons and disks \cite{Wood1990}. Remarkably, both plots are parallel 
to the line $\varphi = \psi$ with an offset $\varphi_0$:  
\begin{equation}
\varphi \simeq \varphi_0 + \psi. 
\end{equation}
The offset $\varphi_0$  is the friction angle at zero dilatancy. 
We have $\varphi_0 \simeq 0.12 $ 
for disks and   $\varphi_0 \simeq 0.3 $ for polygons. This observation is in 
agreement with the  arguments of Taylor \cite{Wood1990,Radjai2004} based on 
energy balance and recently revisited also 
in the case of cohesive granular media \cite{Taboada2006}.    
The higher level of $\varphi$  for the polygon packing reflects the 
organization of the microstructure and the features of force 
transmission for each particle shape. 
This point is considered in more detail in the following section.  

\section{Granular texture}
\label{texture}

The granular texture, i.e. the organization of the particles and their contacts in space, 
is basically controlled by steric exclusions between the particles and 
force balance conditions \cite{Troadec2002}. 
The texture can be described in terms of various statistical descriptors pertaining to 
the force-bearing network of particles. At the lowest order, the compactness of the structure 
can be described in terms of both the solid fraction $\rho$ and the 
coordination number $z$. The connectivity of the network can further be characterized by the 
fraction $P(c)$ of particles having exactly $c$ contact neighbors. These are 
scalar parameters or functions. 
At higher orders, the anisotropy of the texture is described by different ``fabric tensors". We 
consider here these geometrical descriptors in order to identify the signature of 
particle shape.         

\subsection{Connectivity}

\begin{figure}
\includegraphics[width=8cm]{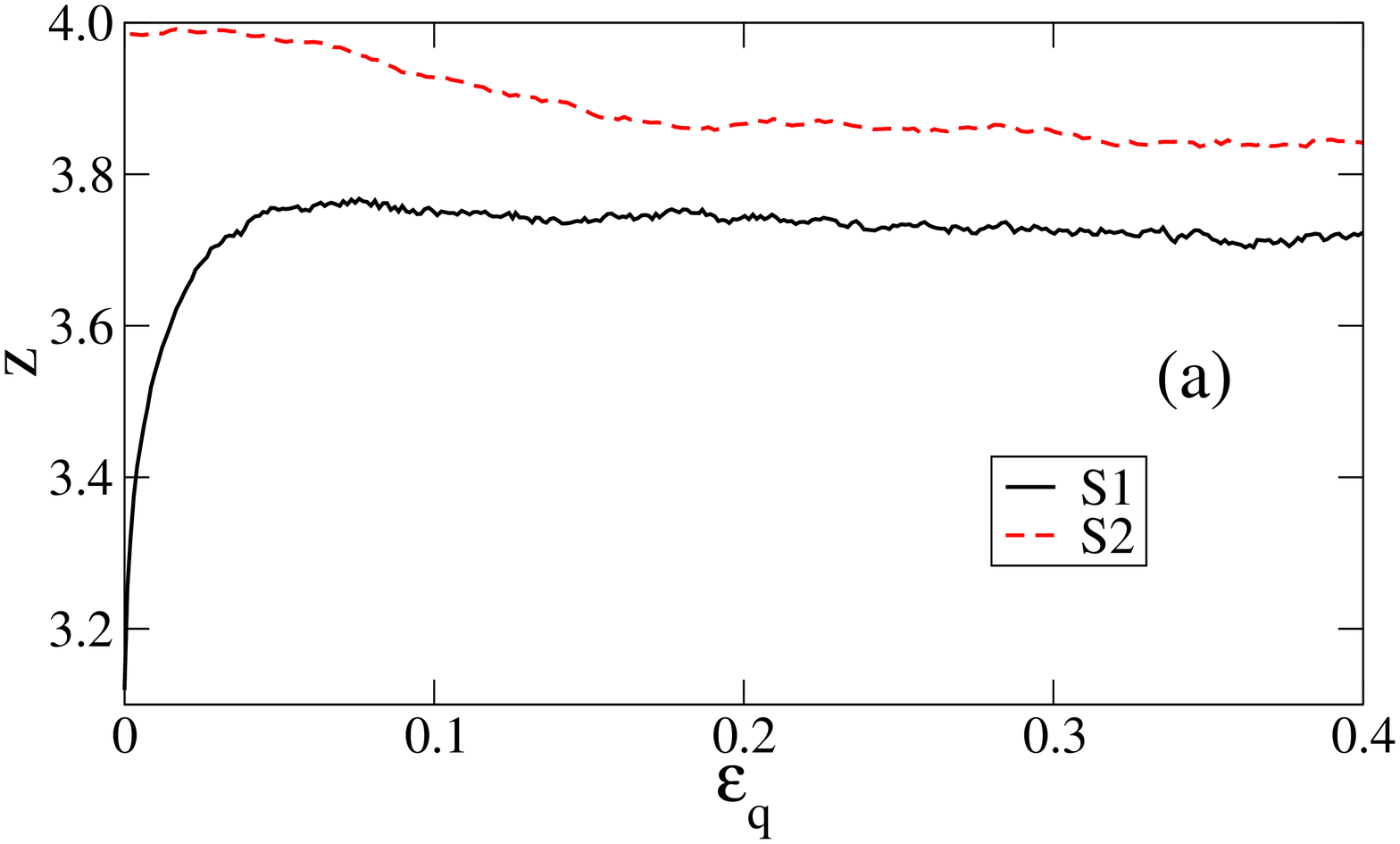}
\includegraphics[width=8cm]{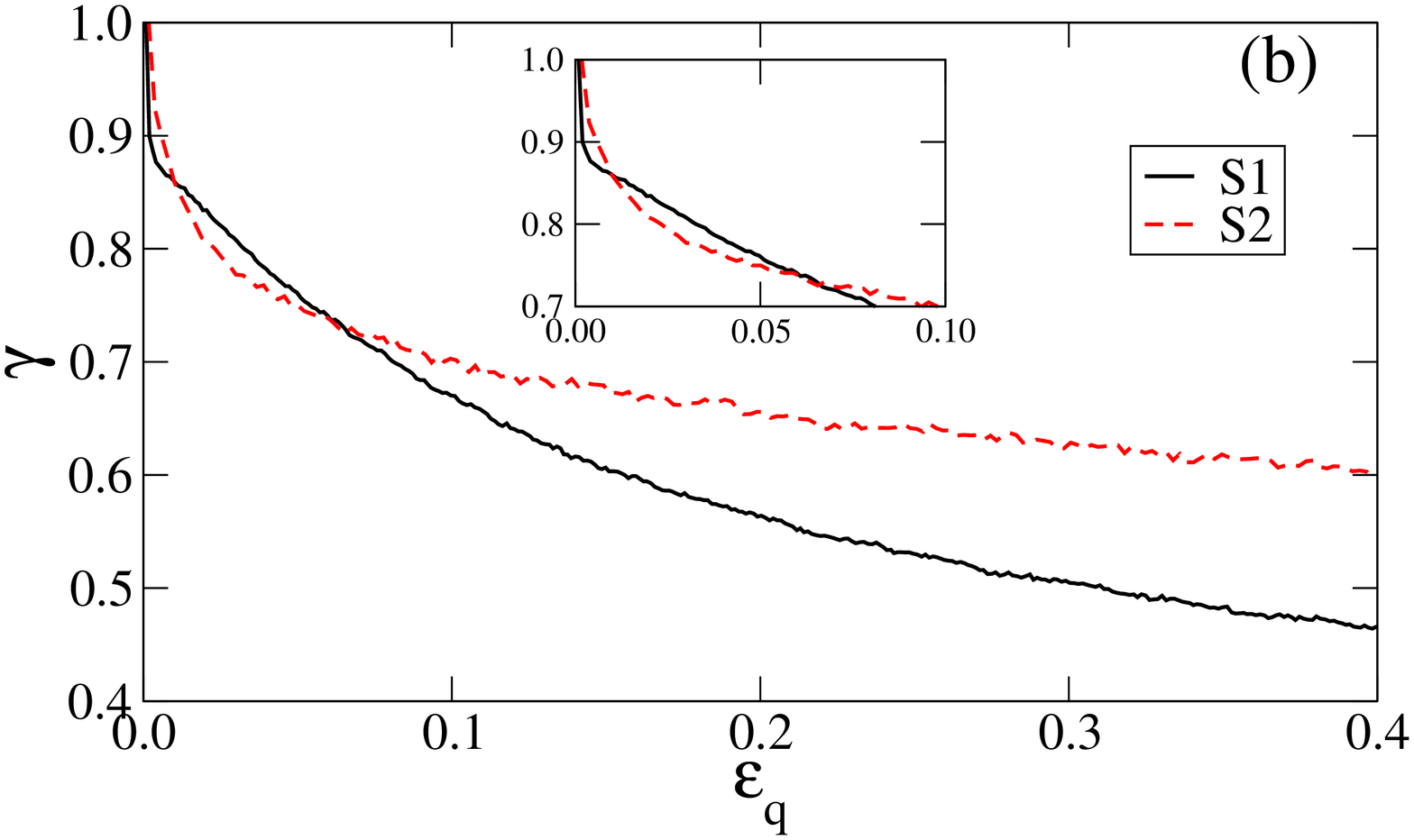}
\caption{The coordination number $z$ (a) and the fraction $\gamma$ of 
persistent contacts (b) as a function of cumulative shear strain $\varepsilon_q$ 
for the samples S1 and S2.   \label{fig06}}
\end{figure}

The connectivity of the particles by force-bearing contacts is described at the lowest order by the 
average number $z$ of contact neighbors per particle. The particles with no force-bearing 
contact are thus removed from the statistics. Note also that each double (edge-to-edge) contact 
for the polygons is counted once  although  double contacts 
are treated as two point contacts belonging to the contact 
segment (see section \ref{procedure}). 
Fig. \ref{fig06}a displays the evolution of $z$ for the pentagon packing (S1) and 
the disk packing (S2) as a function of $\varepsilon_q$. The coordination number  
evolves to  a steady-state value in both samples that is higher for S2 ($\simeq 3.85$) 
than for S1 ($\simeq 3.75$). The difference is, however, much less important than in the 
initial configuration ($\simeq 3.95$ for S2 compared to $\simeq 3.20$ for S1) 
prepared by means of isotropic compaction. 

It is also interesting to compare the two samples in terms of  ``contact lifetimes". 
Let us consider a reference configuration, e.g. the initial state of each sample. 
We follow the history of each contact listed in this state. 
In particular, we define $\gamma$ as the fraction of  
persistent contacts of the initial list. During deformation, 
$\gamma$ declines from $1$ to $0$ as an increasing number of initial contacts are lost due 
to particle rearrangements. 
Fig. \ref{fig06}b shows $gamma$  as a function of $\varepsilon_q$ for S1 and S2. 
We see that, following a rapid initial falloff,  
$\gamma$ decreases slowly in both samples but the rate of contact loss is globally higher 
for polygons than disks. We remark that  even at  
$\varepsilon_q = 0.4$, the contact list is renewed by only $50\%$. 

\begin{figure}
\includegraphics[width=8cm]{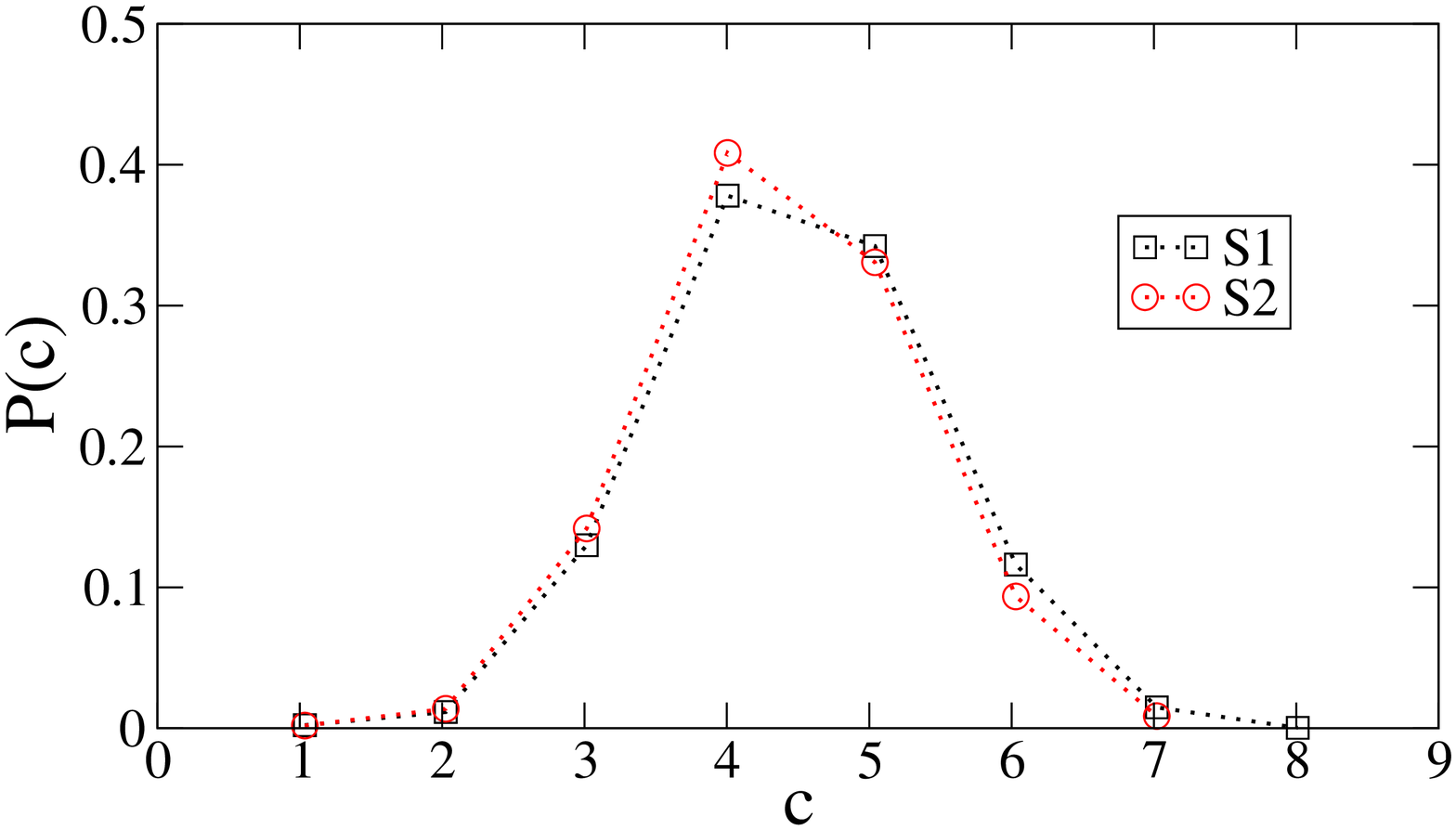}
\caption{Connectivity diagram for the samples S1 and S2 expressing the 
fraction $P(c)$ of particles with exactly $c$ contacts in the residual state.      \label{fig07}}
\end{figure}

The connectivity $P(c)$ of the particles is plotted in Fig. \ref{fig07} for S1 and S2 
at $\varepsilon_q = 0.3$. Interestingly, the 
two plots are nearly identical with a peak for $c=4$. In both samples, the fraction of particles with 
5 contacts is larger than that with 3 contacts. This shows that the connectivity does not 
reflect the difference in texture between the two packings although a qualitative difference exists 
as we shall see below by considering fabric  anisotropy and force transmission.        

\subsection{Fabric anisotropy}

The shear strength of dry granular materials is generally attributed to  
the buildup of an anisotropic structure during shear due to 
friction between the particles and as a result of steric effects depending on  
 particle shape \cite{Oda1980,Cambou1993,Radjai2004a}. 
Several methods have been used  
to quantify the fabric (structural) anisotropy of granular 
materials \cite{Satake1982,Rothenburg1989,Oda1999}. 
A common approach is to  consider 
the probability distribution $P({\bm n})$ of the contact normals $\bm n$ which are 
generically nonuniform. 
In two dimensions, the unit vector $\bm n$ is described by a 
single angle $\theta$, the orientation of the contact normal. 
The probability density function $P_\theta (\theta)$ of contact normals 
provides a detailed statistical information about the fabric. It is $\pi$-periodic 
in the absence of an intrinsic polarity for  $\bm n$. 

Most structural information is generally condensed in the 
second moment of $P_\theta$, 
called {\em fabric tensor}  \cite{Satake1982}:  
\begin{equation}
F_{\alpha \beta} = 
\frac{1}{\pi} \int_0^\pi  n_\alpha (\theta) n_\beta (\theta)  P_\theta(\theta) d\theta 
\equiv \frac{1}{N_c} \sum_{c\in V} n_\alpha^c n_\beta^c, 
\label{eq:F}
\end{equation}
where $\alpha$ and  $\beta$ design the components in a reference frame and 
$N_c$ is the 
total number of contacts in  the control volume $V$.  
By definition, $tr ({\bm F}) = 1$. The 
anisotropy of the contact network is given the difference between the principal values 
$F_1$ and $F_2$. We define  the fabric anisotropy $a$ by 
\begin{equation}  
a = 2 (F_1 - F_2).
\end{equation}
For fix coordinates, with the x-axis pointing along $\theta'$, we 
define also a "signed anisotropy" $a'$ by
\begin{equation}  
a' = 2 (F_1 - F_2) \cos 2(\theta_F - \theta'),
\label{eq:a'}
\end{equation}
where $\theta_F$ is the major principal direction of the fabric tensor. For $\theta' = \theta_F$, we have 
$a' = a$. The signed anisotropy  corresponds to the second term of the  
Fourier expansion of $P_\theta(\theta)$ and 
it is useful whenever the direction of anisotropy is not constant. 
  
\begin{figure}
\includegraphics[width=8cm]{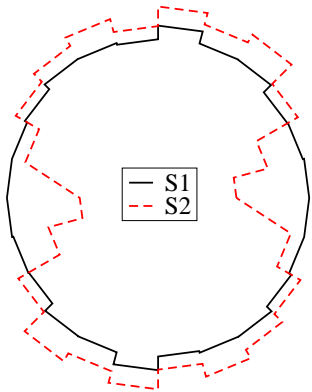}
\caption{Polar representation of the probability density function 
$P_\theta$ of the contact normal directions $\theta$ 
for the samples S1 and S2 in the residual state.     \label{fig08}}
\end{figure}

Figure \ref{fig08} displays a polar representation of 
$P_\theta(\theta)$ for the samples S1 and S2 at $\varepsilon_q = 0.3$. We observe 
a nearly isotropic distribution for the pentagon packing in spite of shearing whereas the 
disk packing is markedly anisotropic.  This is a surprising observation in view of 
the higher shear strength of the pentagon packing (Fig. \ref{fig03}). It 
is also counterintuitive as one expects that double contacts should allow a polygon packing to 
build more easily an anisotropic structure.   

\begin{figure}
\includegraphics[width=8cm]{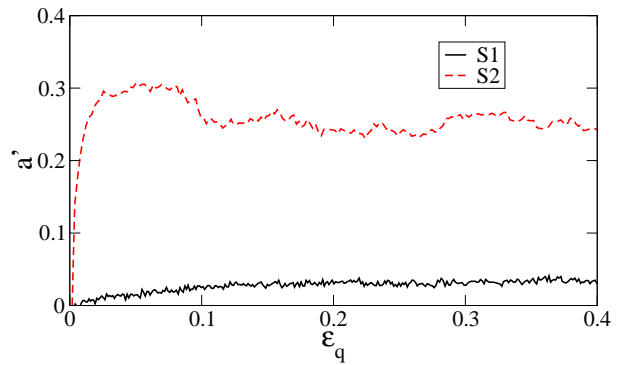}
\caption{Evolution of the anisotropy $a'$ with  cumulative shear strain $\varepsilon_q$
for the samples S1 and S2.    \label{fig09}}
\end{figure}

The evolution of $a'$ is shown in Fig. \ref{fig09} as a function of $\varepsilon_q$ for 
S1 and S2.  The privileged direction of the contacts, corresponding to $\theta_F$, is vertical in both packings.
In both cases, $a'$ increases from 0 (as a result of the initial isotropic compression)
to its largest value in the residual state.
The anisotropy stays quite weak  in the pentagon packing whereas the disk packing 
is marked by a much larger anisotropy, increasing to $\simeq 0.3$ 
and then relaxing to a slightly lower value in the residual state. 
As we shall see below, the low anisotropy of the pentagon packing results from a particular 
organization of the force network in correlation with the orientations of simple and double contacts in 
the packing (section \ref{sd}).  We will also show that the large shear strength of the pentagon 
packing is a consequence of a strong force anisotropy in this packing (see next section).    

\section{Force transmission}
\label{force}

In this section, we analyze the anisotropy and inhomogeneity 
of force networks in the packings of pentagons and disks. 
This leads us to consider the contributions of force and texture anisotropies 
to average shear stresses.     

\subsection{Force anisotropy}

The angular distribution of contact forces in a granular packing can be represented by the 
average force ${\langle \bm f \rangle}({\bm n})$ as a function of the contact normal direction $\bm n$. 
We  distinguish the average normal force $\langle f_n \rangle$ from the average 
tangential force $\langle f_t \rangle$ formally defined  by  \cite{Rothenburg1989}
\begin{equation}
\left\{
\begin{array}{lcl}
\langle f_n \rangle (\theta) &=& \frac{1}{N_c(\theta)} \sum\limits_{c \in {\cal S}(\theta)} f_n^c, \\
\langle f_t \rangle (\theta) &=& \frac{1}{N_c(\theta)} \sum\limits_{c \in {\cal S}(\theta)} f_t^c, 
\end{array}
\right.
\end{equation}
where $f_n^c$ and $f_t^c$ are the normal and tangential forces, respectively, acting at the 
contact $c$ (according to a sign convention attributing positive values to the normal forces),  
${\cal S}(\theta)$ is the set of contacts with 
direction $\theta \in [\theta- \Delta\theta/2, \theta+ \Delta\theta/2]$ for angle 
increments $\Delta\theta$, and $N_c(\theta)$ is the number of contacts in ${\cal S}(\theta)$.     

By definition, the two functions $\langle f_n \rangle$ and $\langle f_t \rangle$ are $\pi$-periodic. 
After sufficiently long  monotonous shearing, these functions 
can be approximated by their Fourier expansions 
truncated beyond the second term \cite{Rothenburg1989,Radjai2004a}:
\begin{equation}
\left\{
\begin{array}{lcl}
\langle f_n \rangle (\theta) &=& \langle f \rangle  \{ 1 + a_n \cos 2(\theta - \theta_n) \}  \\
\langle f_t \rangle (\theta) &=&  \langle f \rangle  a_t \sin 2(\theta - \theta_t) , 
\end{array}
\right.
\label{eqn:fnft}
\end{equation}  
where $\langle f \rangle$ is the average force, $a_n$ and $a_t$ represent the anisotropies of the 
normal and tangential forces, respectively, and  $\theta_n$ and $\theta_t$ are their 
privileged directions.  

\begin{figure}
\includegraphics[width=4cm]{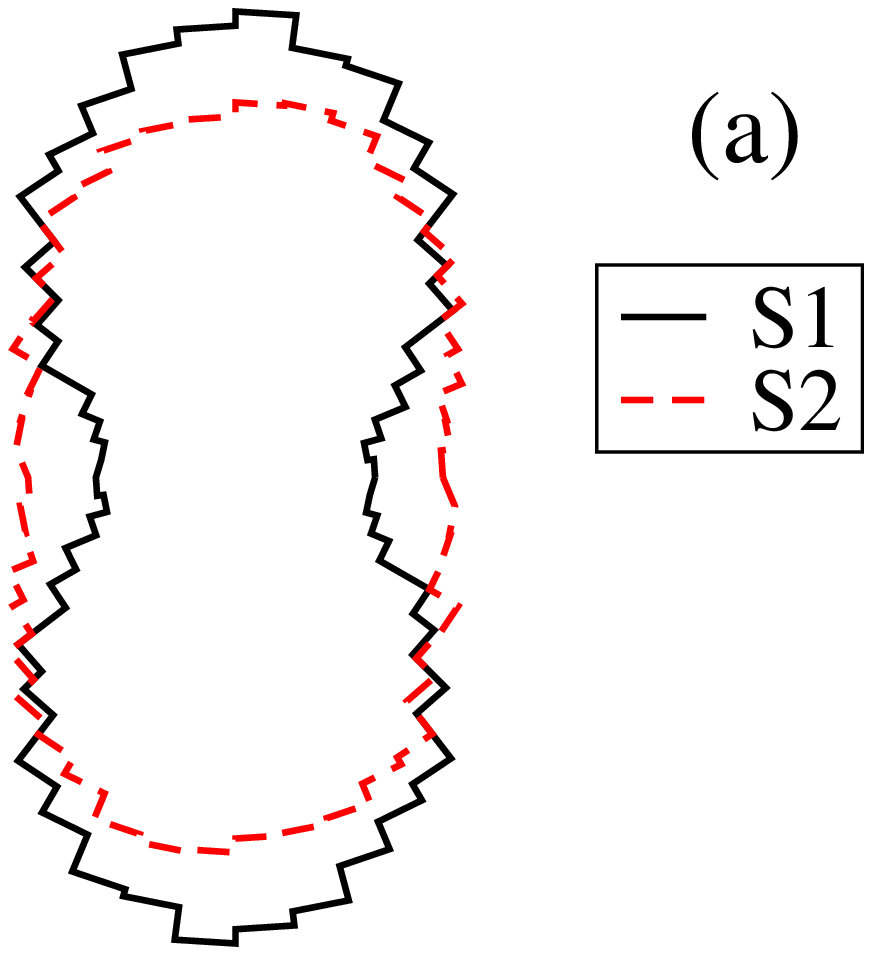}
\includegraphics[width=5cm]{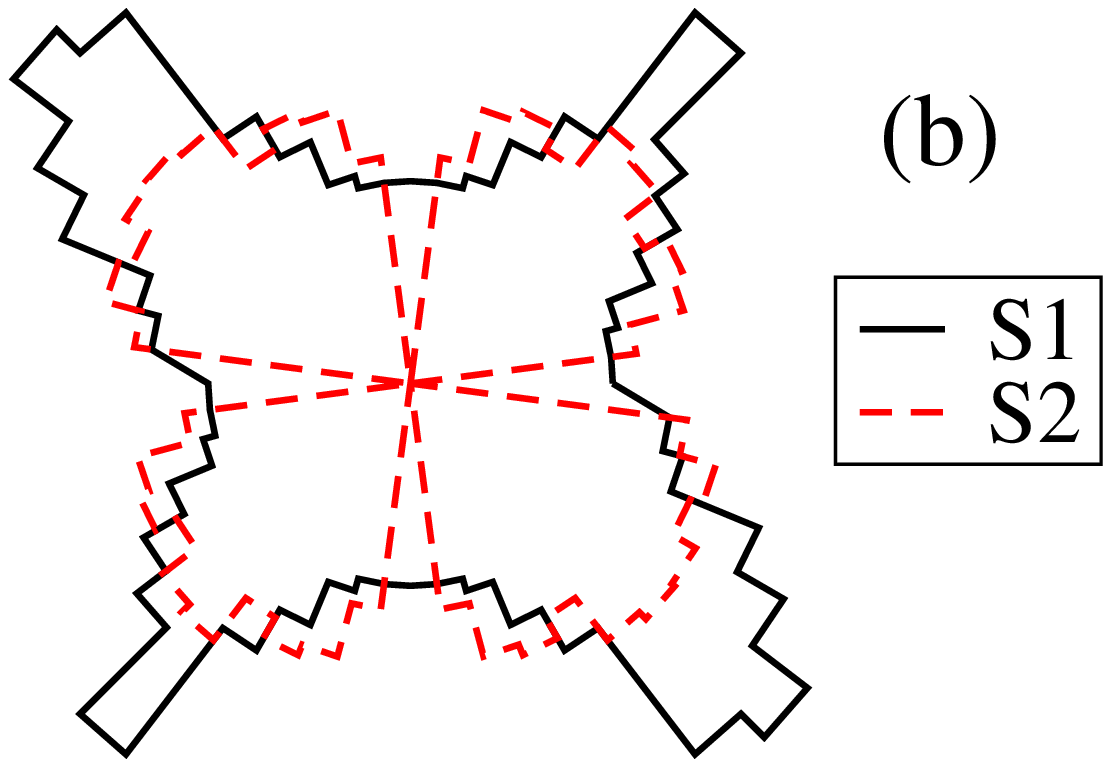}
\caption{Polar representation of the angle-averaged normal (a) and tangential (b) forces 
$ \langle f_n \rangle (\theta)$ and 
$\langle f_t \rangle (\theta)$ for the samples S1 and S2 in the residual state.   \label{fig10}}
\end{figure}

In Fig. \ref{fig10},  the functions $ \langle f_n \rangle (\theta)$ and 
$\langle f_t \rangle (\theta)$ are displayed in polar coordinates at $\varepsilon_q = 0.3$. 
The pentagon and disk packings show pronounced force anisotropy with a 
stronger anisotropy in the case of pentagons both for normal and tangential forces. These 
plots can be fitted by harmonic functions [Eq. (\ref{eqn:fnft})] in order to estimate the force anisotropies 
 $a_n$ and $a_t$. However, it is more convenient to estimate the 
 anisotropies through  the following ``force tensors'': 
\begin{equation}
\left\{
\begin{array}{lcl}
H^{(n)}_{\alpha \beta} &=& 
 \int\limits_{0}^\pi  
\langle f_n \rangle(\theta)  n_\alpha  n_\beta d\theta ,  \\ 
H^{(t)}_{\alpha \beta} &=& 
 \int\limits_{0}^\pi  
\langle f_t \rangle(\theta)  n_\alpha  n_\beta d\theta.  \\ 
\end{array}
\right.
\label{eqn:HnHt}
\end{equation}      
It is easy to see that $tr({\bm H}^{(n)}) = tr({\bm H}^{(t)}) = \langle f \rangle$, and 
by identification with (\ref{eqn:fnft}) we have  
\begin{equation}
\left\{
\begin{array}{lcl}
a_n &=& 2 \frac{H^{(n)}_1  - H^{(n)}_2}{H^{(n)}_1  + H^{(n)}_2}, \\
a_t &=& 2 \frac{H^{(t)}_1  - H^{(t)}_2}  {H^{(t)}_1  + H^{(t)}_2}, \\
\end{array}
\right.
\label{eqn:anat}
\end{equation}  
where the subscripts $1$ and $2$ refer to the principal values of the tensors.  

\begin{figure}
\includegraphics[width=8cm]{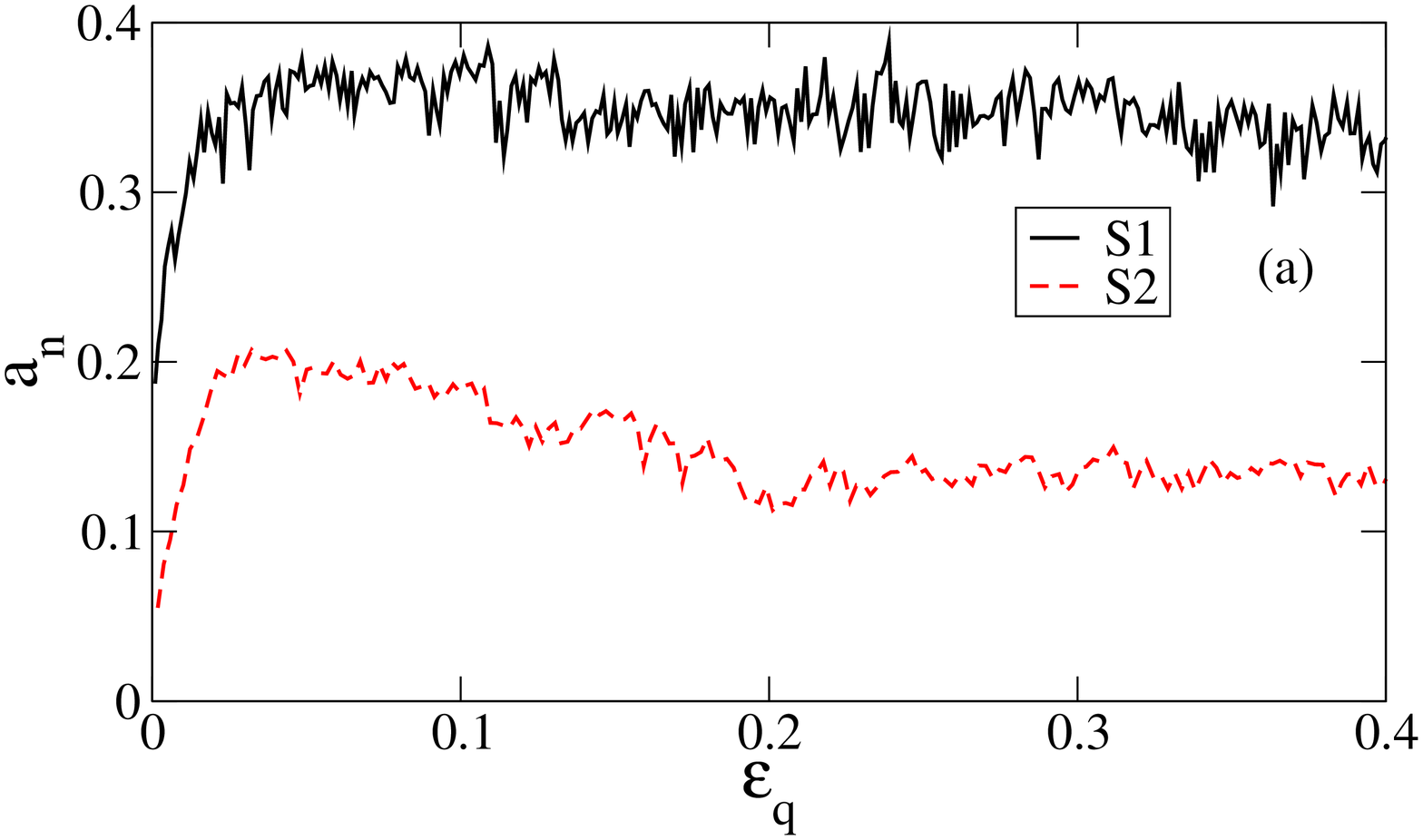}
\includegraphics[width=8cm]{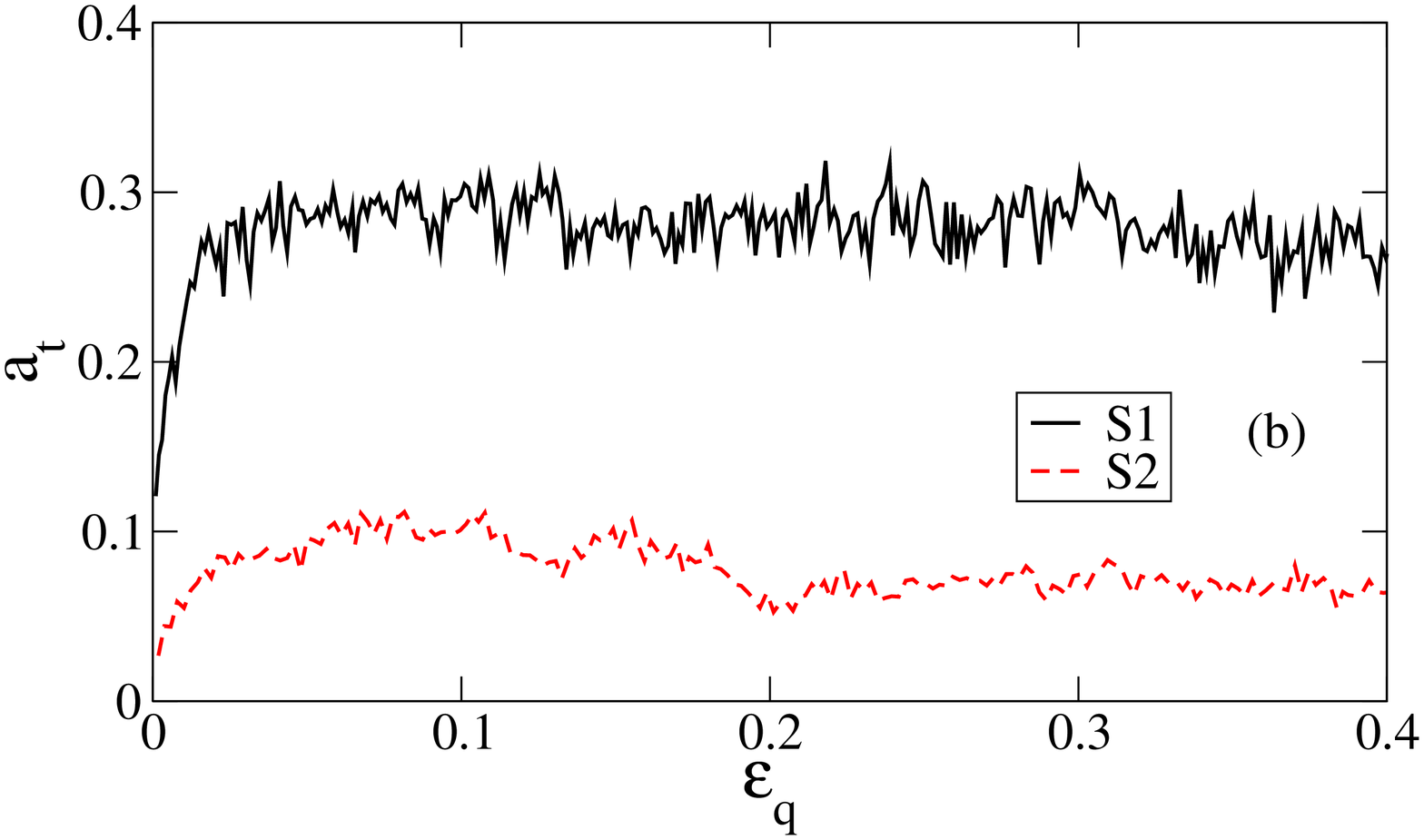}
\caption{Evolution of force anisotropies  $a_n$ (a) and $a_t$ (b) as a function 
of cumulative shear strain  $\varepsilon_q$ in samples S1 and S2.  \label{fig11}}
\end{figure}

Figure \ref{fig11} 
shows the evolution of $a_n$ and $a_t$ with $\varepsilon_q$ in samples S1 and S2. We see that, 
in contrast to fabric anisotropies (Fig. \ref{fig09}), 
the force anisotropies in pentagon packing remain always above those in 
the disk packing. This means that the aptitude of the pentagon packing to develop 
large force anisotropy and strong force chains 
is not solely dependent on the global fabric anisotropy of the system. In section \ref{sd}, we will show that 
the force anisotropy of the pentagon packing stems from the high anisotropy of the sub-network of double 
contacts and strong activation of friction forces. Indeed, due to 
the geometry of the pentagons, i.e. the absence of parallel sides, the strong force chains 
are mostly of zig-zag shape, as observed in Fig. \ref{fig13}b, and the 
stability of such structures requires strong activation of tangential forces. This 
explains, in turn, the large value of $a_t$ for pentagons, very close to $a_n$, whereas in the disk packing 
$a_t$ is nearly half of $a_n$.      

\begin{figure}
\includegraphics[width=8cm]{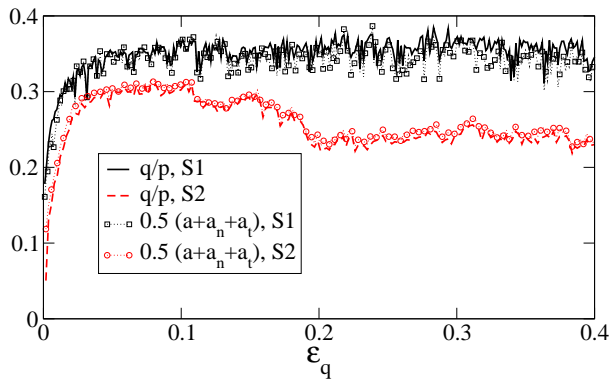}
\caption{Evolution of the normalized shear stress  $q/p$ for the samples S1 and S2 
with $\varepsilon_q$ 
together with the corresponding predictions from its expression as a function of 
fabric and force anisotropies [Eq. (\ref{eq:qa})].     \label{fig12}}
\end{figure}

The  anisotropies $a$, $a_n$ and $a_t$ are interesting descriptors of granular 
microstructure and force transmission as they underlie the 
shear stress. Indeed,     
it can be shown that the general expression of the stress tensor Eq. (\ref{eq:M}) leads to the 
following simple relation \cite{Rothenburg1989,Radjai2004a}:
\begin{equation}     
\frac{q}{p} \simeq \frac{1}{2} (a+a_n+a_t),
\label{eq:qa}
\end{equation}
where the cross products between the anisotropies have been neglected 
and it has been assumed that the stress tensor is coaxial with the fabric tensor Eq. (\ref{eq:F}) 
and the force tensors Eq. (\ref{eqn:HnHt}). 
Fig. \ref{fig12} shows that Eq. (\ref{eq:qa}) holds quite well both for pentagons and 
disks. 

A remarkable consequence of Eq. (\ref{eq:qa}) is to reveal 
the distinct origins of shear stress in pentagon and disk packings. 
The fabric anisotropy provides a major contribution to  shear stress in the disk packing 
(Fig. \ref{fig09}) whereas 
the force anisotropies are more important for shear stress in 
the pentagon packing (Fig. \ref{fig11}). In this way, in spite of the weak fabric anisotropy $a$, 
the larger force anisotropies $a_n$ and $a_t$ allow 
the pentagon packing  to reach higher levels of $q/p$ compared to the disk packing.

\subsection{Force distributions}
The strong inhomogeneity of contact forces  is a well-known feature 
of granular media. It has been investigated mostly for 
spherical or cylindrical particles both by experiments and numerical 
simulations \cite{Antony2001,Liu1995a,Majmudar2005,Mueth1998a,Radjai1996,Lovol1999,
Silbert2002,Bardenhagen2000}. 
The probability density function (pdf) of normal forces is characterized by 
two features which seem to be specific to granular media: 1) The pdf is roughly a decreasing 
exponential function for forces above the mean, 2) In the range of weak forces below the mean, 
the pdf does not decline to zero with the force. The relative scatter of data reported by different authors 
for weak forces shows the sensitivity of the pdf in this range to the details 
of the microstructure. But, the common observation that there is a large number of contacts 
transmitting very weak forces, is a straightforward signature of the arching effect. 
From this point of view, one expects that angular particle shape will influence mainly 
the distribution of weak forces by enhancing the arching effect.  
 
\begin{figure}
\includegraphics[width=8cm]{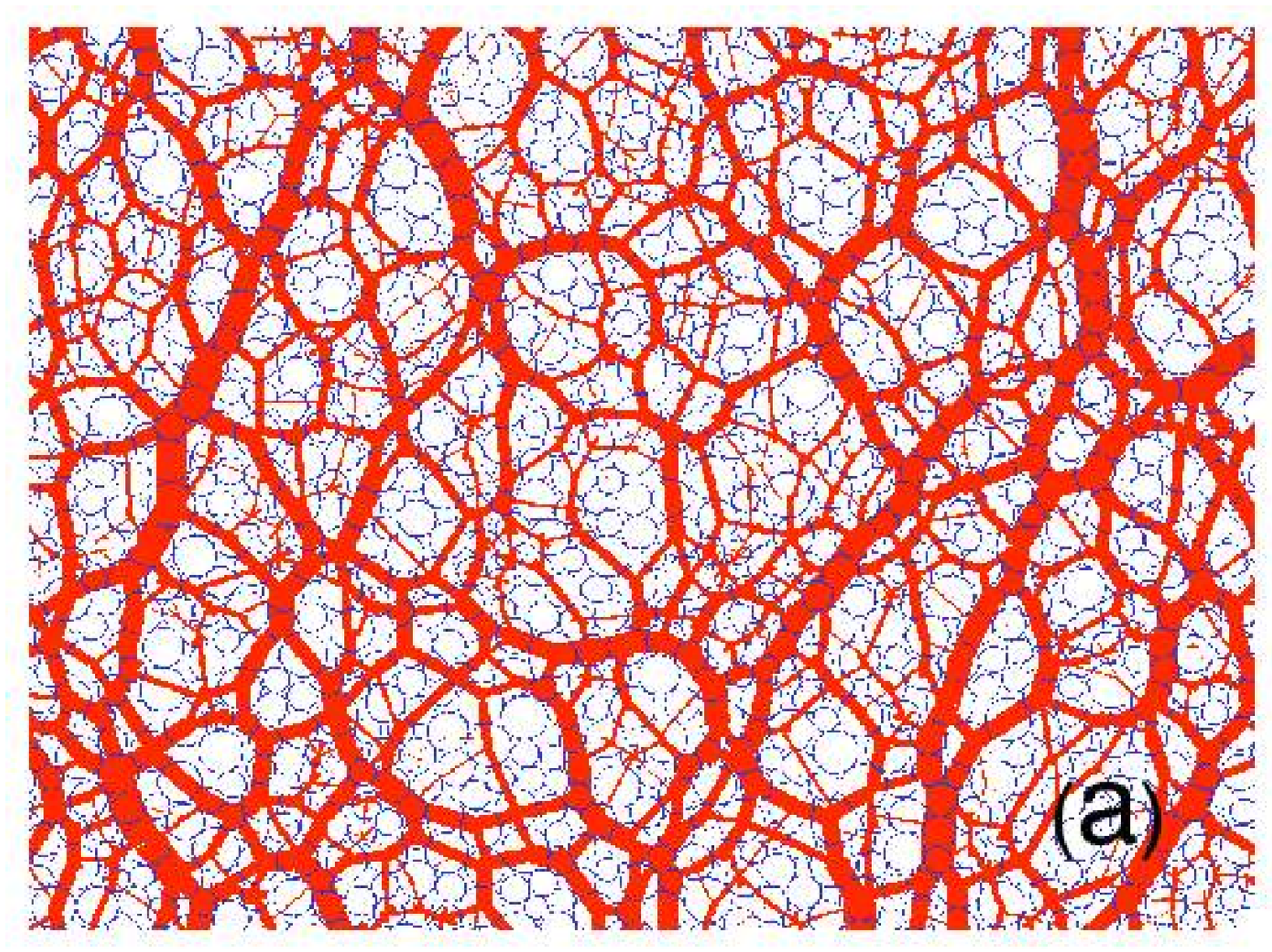}
\includegraphics[width=8cm]{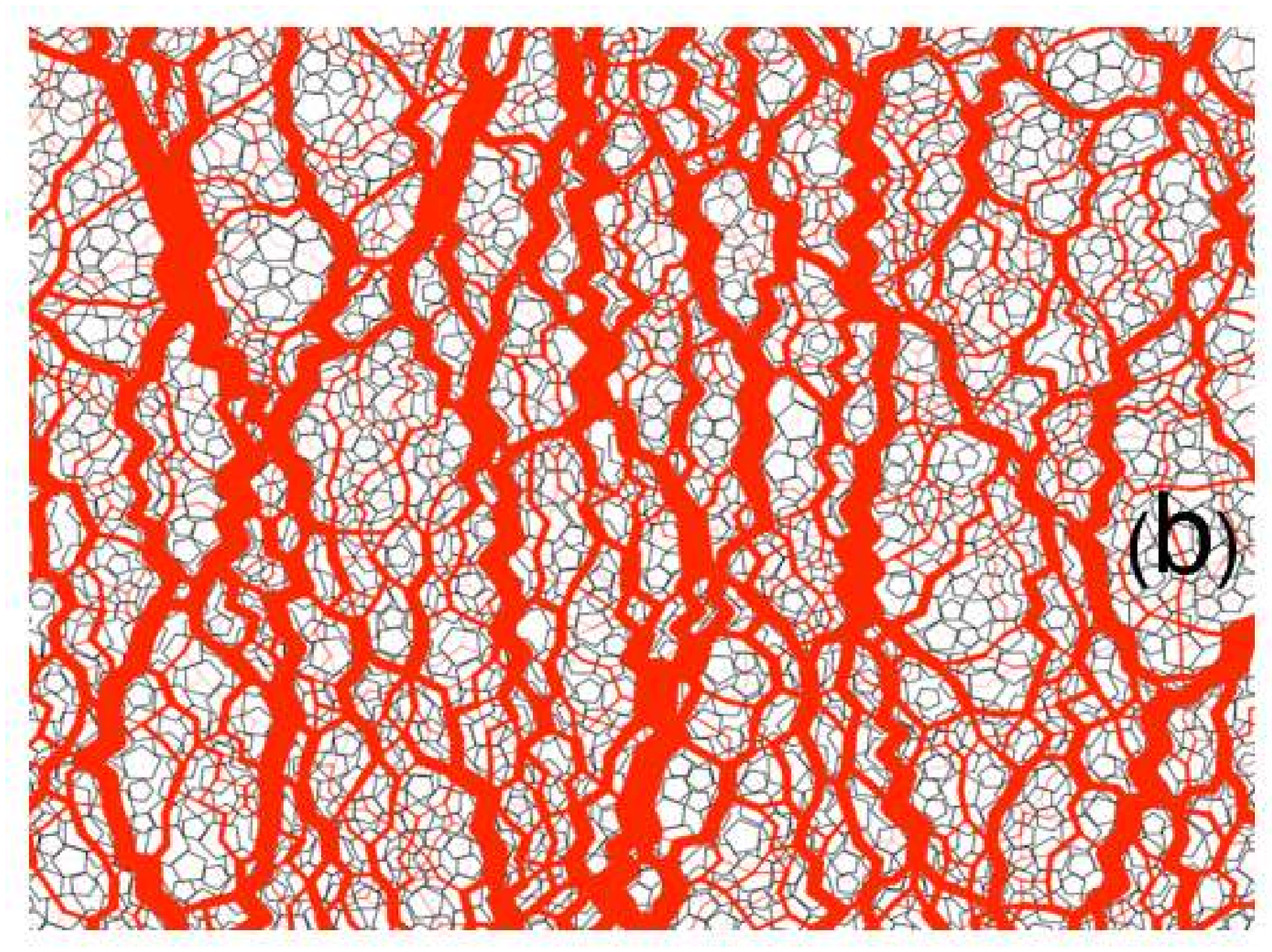}
\caption{(color online) Snapshots of normal forces in samples S2 (a) and S1 (b). Line thickness is proportional 
to the normal force. \label{fig13}}
\end{figure}

\begin{figure}
\includegraphics[width=8cm]{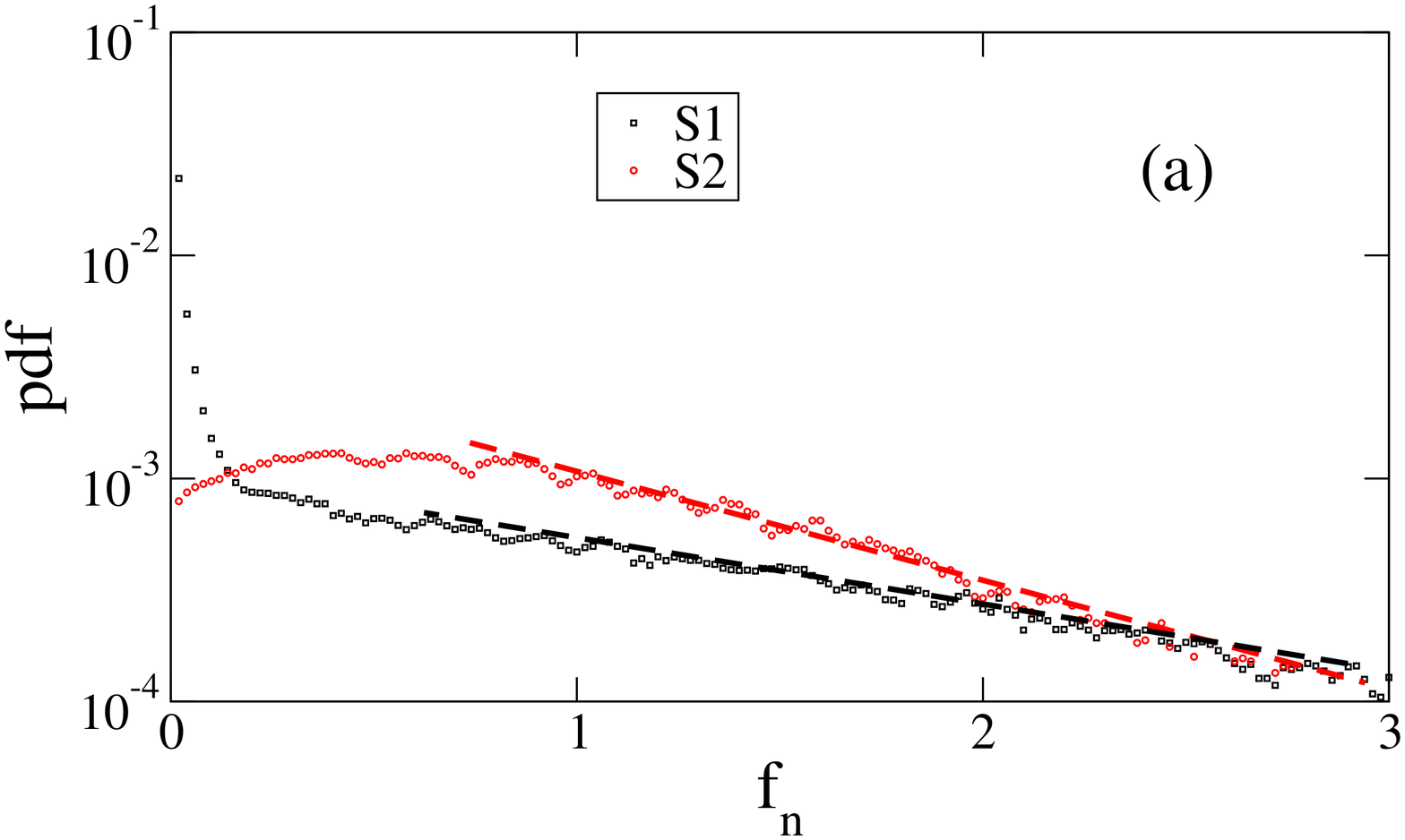}
\includegraphics[width=8cm]{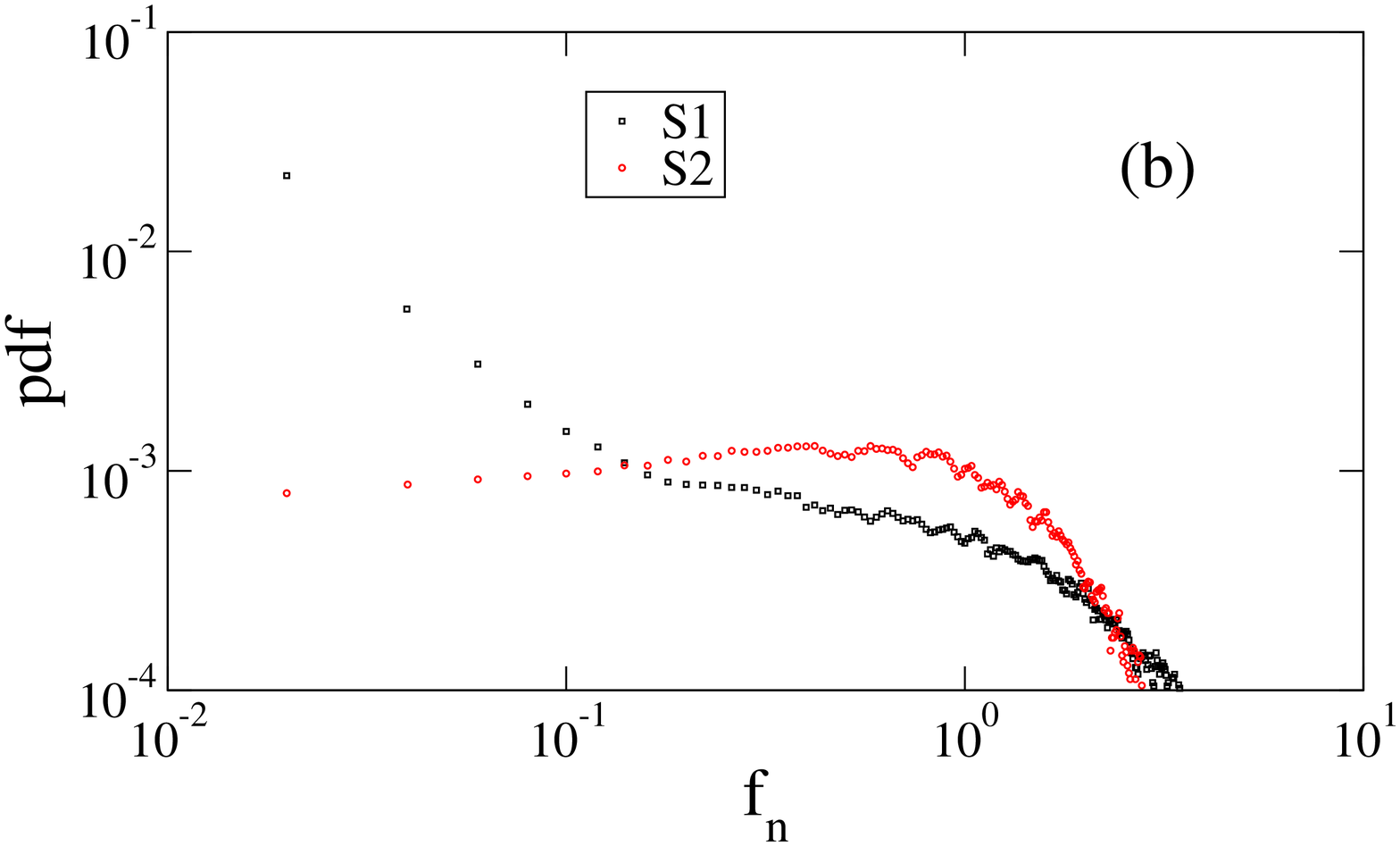}
\caption{Probability density functions of normal forces in samples S1 and S2 
in log-linear (a) and log-log scales (b).    \label{fig14}}
\end{figure}

Figure \ref{fig13} displays maps of normal forces in a portion of 
each of the samples S1 and S2 at a  large cumulative strain. 
We observe the strong anisotropy of normal forces in the pentagon packing 
compared to the disk packing (as discussed in section \ref{force}) as well as the zig-zag form 
of the strong  force chains. The normal force pdf's are shown in Fig. \ref{fig14} in log-linear and log-log scales at 
large strains. The forces are normalized by the mean normal force $\langle f_n \rangle$ in each sample. In both samples, the number of  strong forces (above the mean $\langle f_n \rangle$) falls off 
exponentially:
\begin{equation}
\left\{
\begin{array}{lcl}
f_n \propto e^{- \alpha_1  f_n / \langle f_n \rangle} \;  \; \; \mbox{in S1}, \\
f_n \propto e^{- \alpha_2  f_n / \langle f_n \rangle} \;  \; \; \mbox{in S2}, 
\end{array}
\right.
\label{eqn:strong}
\end{equation}      
with $\alpha_1 \simeq 0.74$ and $\alpha_2 \simeq 1.4$. The smaller value of $\alpha_1$ 
means that the distribution is wider for pentagons compared to disks. The 
distribution is nearly uniform in the whole range of weak forces  
($f_n < \langle f_n \rangle$) in  
S2.  In the pentagon packing S1, we observe a uniform distribution only in the range  
$ 0.1\langle f_n \rangle < f_n < \langle f_n \rangle$. Nearly $30 \%$ of forces are in this range. 
The number of ``very 
weak" forces in S1 in the range  $f_n < 0.1 \langle f_n \rangle$ increases 
faster than a power law as $f_n$ tends to zero. A fraction  $\simeq  30 \%$ of contacts belong to 
this range. The presence of numerous ``very weak'' forces in the pentagon
 packing is a clear signature 
of  enhanced arching effect that can be characterized, as we shall see below, by the 
respective roles of simple and double contacts with respect to force transmission.

\subsection{Bimodal character of stress transmission}
The genuine organization of contact forces in granular media,  
involving strong force chains 
propped by weak forces, was first analyzed   
by Radjai et al. by means of contact dynamics simulations 
for packings of circular and spherical particles \cite{Radjai1998}. 
This analysis proceeds by considering the subset of contacts which carry a force 
below a  cutoff force $\xi$ normalized by the mean force.  This subset is referred to as the 
``$\xi$-network''. The variation of a quantity evaluated for the
``$\xi$-network'' as $\xi$ is varied from $0$ to the maximal
force in the system, provides its correlation with the 
contact force. Here, we apply this same approach to S1 and S2 samples 
for the stress ratio $q(\xi)/p$, defined as stress deviator $q(\xi)$ (normalized by the total pressure $p$ of the sample) in the $\xi$-network,  
and for $a(\xi)$, defined as the fabric anisotropy in the $\xi$-network. 

\begin{figure}
\includegraphics[width=8cm]{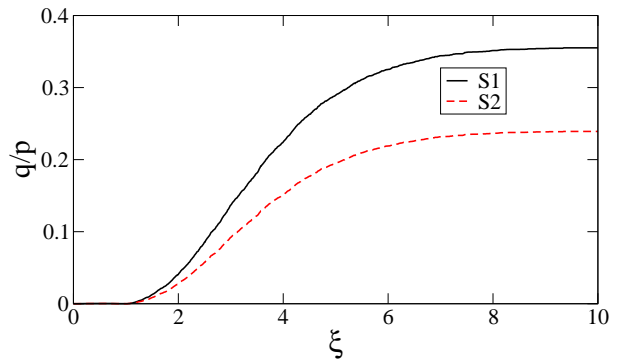}
\caption{Partial shear stress $q(\xi)/p$ as a function of force cutoff $\xi$ (normalized 
by the mean force) for the samples 
S1 and S2 in the residual state.      \label{fig15}}
\end{figure}

\begin{figure}
\includegraphics[width=8cm]{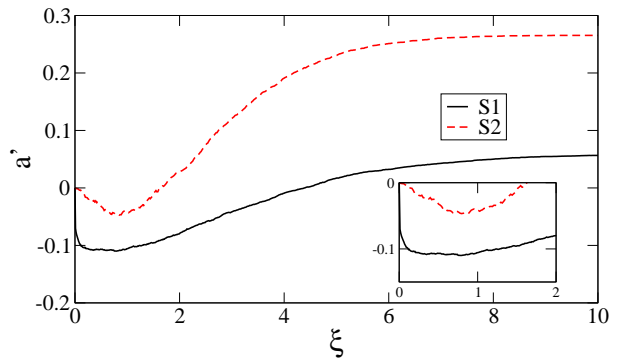}
\caption{Partial fabric anisotropy $a'(\xi)$ as a function of 
force cutoff $\xi$ (normalized 
by the mean force) in the samples S1 and S2.      \label{fig16}}
\end{figure}

The plot of  $q(\xi)/p$ is shown in Fig. \ref{fig15} for S1 and S2 in the residual state. 
In both samples, the stress deviator is nearly zero for $\xi < 1$, i.e. for the normal forces 
below the average force. This means that the  shear stress is almost totally sustained 
by the ``strong" contact network $\xi > 1$ for the pentagon packing as well as 
for the disk packing. Fig. \ref{fig16} shows the fabric anisotropy $a'(\xi)$ as a function of 
$\xi$ in the samples S1 and S2. By definition,     
a positive value of $a'$  corresponds to 
the principal stress direction  whereas a negative value 
corresponds to the orthogonal direction.
 We see that the direction of anisotropy is orthogonal to the principal stress direction 
 ($a' < 0$) for weak forces (small $\xi$). This ``orthogonal" anisotropy of the 
 weak forces is more important in the pentagon packing compared to the disk packing, and,   
as shown in the inset to Fig. \ref{fig16}, it is mainly due to ``very weak" forces. 
 When $\xi$ is increased beyond 
$\langle f_n \rangle$, $a'$ becomes less negative and 
finally changes sign, showing that 
the strong contacts  are
preferentially parallel to the principal axis. 
These strong contacts are less than 40\% of all contacts, but their positive
contribution to $a'$ overcompensates the negative contribution weak contacts.  
For large $\xi$, the partial anisotropy  approaches the fabric anisotropy of the whole
system.

\begin{figure}
\includegraphics[width=8cm]{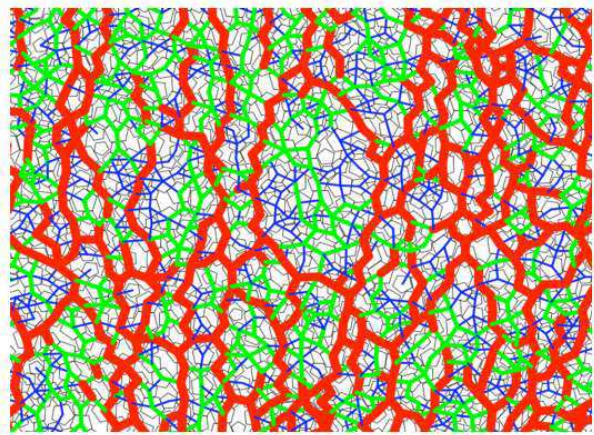}
\caption{(color online) Tricolor map of the contact network composed of very weak (blue), intermediate (green) and 
strong (red) contacts in the pentagon  packing.      \label{fig17}}
\end{figure}

These data demonstrate the bimodal character of stress transmission also 
in the pentagon packing in spite of a very different particle geometry. The mean force 
plays a particular role in differentiating strong contacts  from weak contacts. 
However, the force pdf's (Fig. \ref{fig14}) and the anisotropy of weak 
forces (Fig. \ref{fig16}) provide also evidence for the existence of a class of very weak forces, 
corresponding approximately to the range  $f_n < 0.1 \langle f_n \rangle$, within 
the weak network. This class is strongly anisotropic with a 
privileged direction which is orthogonal 
to the major principal stress direction, and the corresponding 
force pdf diverges as the force tends to zero. Fig. \ref{fig17} displays a 
tricolor map of the contact network representing  
very weak, intermediate ($ 0.1\langle f_n \rangle < f_n < \langle f_n \rangle$) and 
strong contacts 
in the pentagon packing. Large cells of strong contacts are composed of zig-zag 
chains. The anisotropy of strong contacts is reflected in the elongated shape of these cells 
along the major principal stress direction. Both intermediate and very weak 
forces prop these cells.

\section{Simple versus double contacts}
\label{sd}

In this section, we focus on the organization of simple and double contacts 
in the pentagon packing. The double contacts, i.e. the side-sharing polygons, 
are generally assumed to be at the source of the higher strength of polygon 
packings. For the texture, we would like also to investigate the proportions of 
simple and double contacts and their  
respective contributions to the overall anisotropy of the pentagon packing. 
It is also  
important to identify  the role of  double contacts  in force transmission. 

\begin{figure}
\includegraphics[width=8cm]{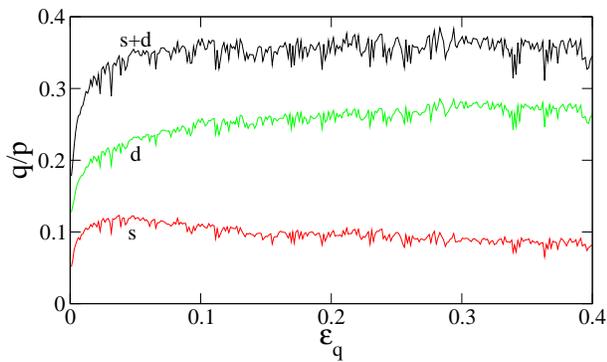}
\caption{Normalized shear stress $q/p$ for simple (s) and double (d) contacts, as 
well as for all contacts (s+d),    
as a function of cumulative shear strain $\varepsilon_q$ in the pentagon packing.   \label{fig18}}
\end{figure}

\begin{figure}
\includegraphics[width=8cm]{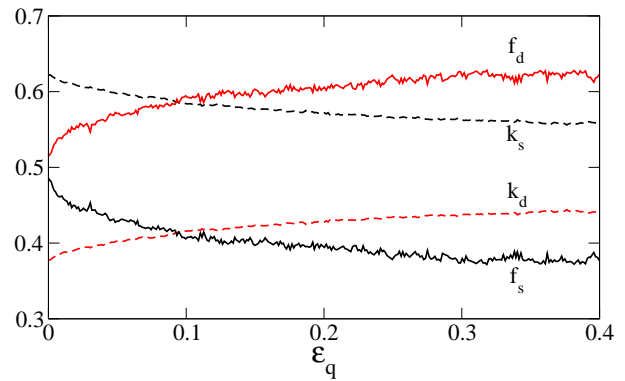}
\caption{Proportions $k_s$ and $k_d$ of simple and double contacts, and the corresponding    
relative force averages $f_s$ 
and $f_d$, as a function  of cumulative shear strain $\varepsilon_q$.    \label{fig19}}
\end{figure}

The general expression [Eq. \ref{eq:M}] of the stress tensor $\bm \sigma$ allows us  
to perform a unique additive decomposition of the stress into two parts: 
\begin{equation}
{\bm \sigma} = {\bm \sigma}_s + {\bm \sigma}_d,
\label{eqn:sigmasd}
\end{equation}
where  ${\bm \sigma}_s$ is obtained from the expression (\ref{eq:M}) by  
restricting the summation  to simple contacts, and  ${\bm \sigma}_d$ is the 
complementary tensor involving only double contacts.  The respective 
stress deviators $q_s$ and $q_d$ normalized by the mean stress $p$ are shown in Fig. \ref{fig18} 
as a function of strain $\varepsilon_q$. The strength $q_d/p$ of double contacts varies from two to three times  
that of simple contacts during shear deformation 
of the pentagon packing. The proportions $k_s$ and $k_d$ of simple and double contacts are shown in Fig. \ref{fig19}
 as a function  of $\varepsilon_q$. The same figure displays 
the relative force averages $f_s = k_s \langle f_n \rangle_d / \langle f_n \rangle$ 
and $f_d= k_d \langle f_n \rangle_s / \langle f_n \rangle$, where 
$\langle f_n \rangle_s$ and $\langle f_n \rangle_d$ are the mean normal forces of 
simple and double contacts, respectively. We see that $k_d$ 
increases with strain but remains below $k_s$. On the other hand, initially we have  $f_d=f_s=0.5$, reflecting the isotropic state of the packing prepared by 
isotropic compaction. However, $f_d$ increases with shear up to 
$f_d \simeq 1.5 f_s$ in the residual state. This means that    
the larger shear stress carried 
by double contacts in the residual state is due to the larger mean normal force of  
double contacts despite their smaller proportion in the packing.       

\begin{figure}
\includegraphics[width=8cm]{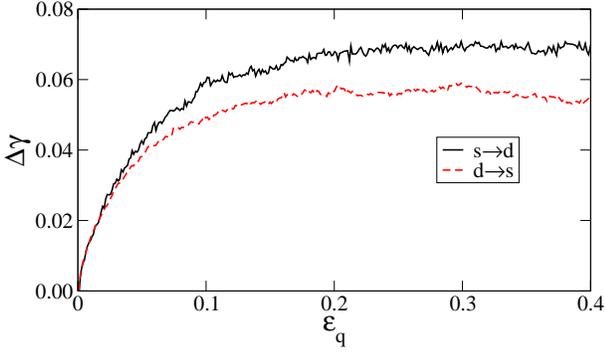}
\caption{Cumulative 
proportion $\Delta \gamma$  of simple contacts turning to 
double (${s\rightarrow d}$) and vice versa (${d\rightarrow s}$).     \label{fig20}}
\end{figure}

The growth of the number of double contacts shown in Fig. \ref{fig19} represents the 
gradual consolidation of the sample. In Fig. \ref{fig20} we plot the cumulative 
proportions $\Delta \gamma_{s\rightarrow d}$ and 
$\Delta \gamma_{d\rightarrow s}$ of simple contacts turning to 
double  and vice versa, respectively. Although transformation between the two contact 
types occurs at each step in both directions $s\rightarrow d$ and $d\rightarrow s$, 
 the consolidation involves on average a net fraction of simple contacts transforming 
into double contacts. 

\begin{figure}
\includegraphics[width=8cm]{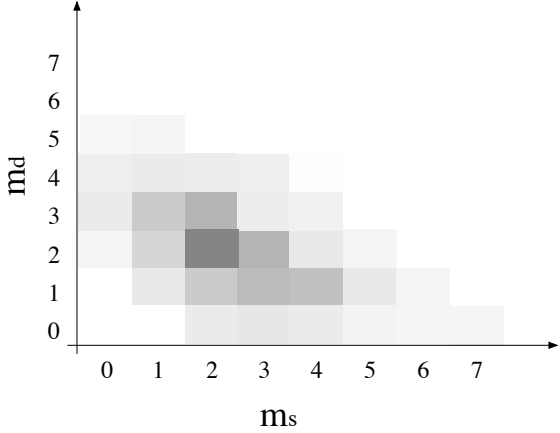}
\caption{Grey level map of the connectivity function $P(m_s,m_d)$ of 
the pentagon packing in the residual state.     \label{fig21}}
\end{figure}

The connectivity of the pentagon packing by simple and 
double contacts can be represented by the proportion 
 $P(m_s,m_d)$ of particles with exactly $m_s$ simple contacts 
 and $m_d$ double contacts. Fig. \ref{fig21} shows a grey level map of this function for the 
 pentagon packing in the residual state. The row $m_d=0$ corresponds 
 to particles with only simple contacts (nearly $2\%$ of the 
 total number of particles) whereas the column$m_s = 0$ 
 represents the particles with only double contacts (nearly $6\%$).  
 On average, a particle has more simple contacts than double contacts but     
 the maximum occurs at $m_s = m_d = 2$.   

\begin{figure}
\includegraphics[width=8cm]{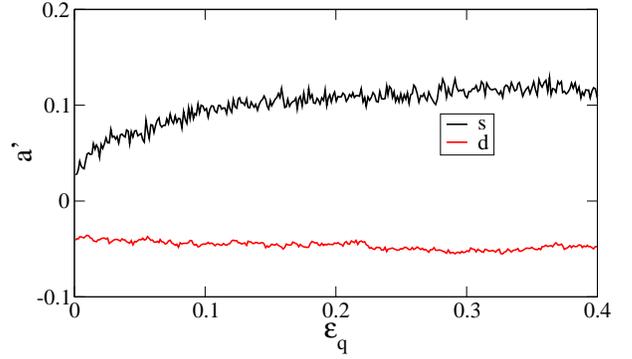}
\caption{The anisotropy $a'$ of simple (s) and double (d) contacts 
as a function of  cumulative shear strain $\varepsilon_q$ in the pentagon packing.    
\label{fig22}}
\end{figure}

We now consider the fabric tensor decomposed in a similar way as 
the stress tensor [Eq. (\ref{eqn:sigmasd})] into two partial tensors:     
\begin{equation}
{\bm F} = {\bm F}_s + {\bm F}_d,
\label{eqn:fabricsd}
\end{equation}
where ${\bm F}_s$ and ${\bm F}_d$ are defined as ${\bm F}$ in Eq. (\ref{eq:F}) 
by simply restricting the 
summation  to simple and double contacts, respectively, and by dividing the sum  
by the total number $N_c$ of contacts. The respective anisotropies $a'_s$ and 
$a'_d$ of simple and double contacts are displayed in Fig. \ref{fig22} as a function of  
$\varepsilon_q$. The interesting observation here is that the 
simple contacts have a negative anisotropy which, according to Eq. (\ref{eq:a'}), 
means that simple contacts are mostly oriented perpendicular to 
the major principal fabric direction $\theta_F$. In other words, most simple contacts 
belong to the weak network. In contrast, the double contacts have an increasing 
positive anisotropy which is larger than the mean anisotropy $a$ of the sample. 
This is consistent with the fact that the double contacts take over larger 
forces and they contribute more to the shear stress than 
simple contacts.         

\begin{figure}
\includegraphics[width=8cm]{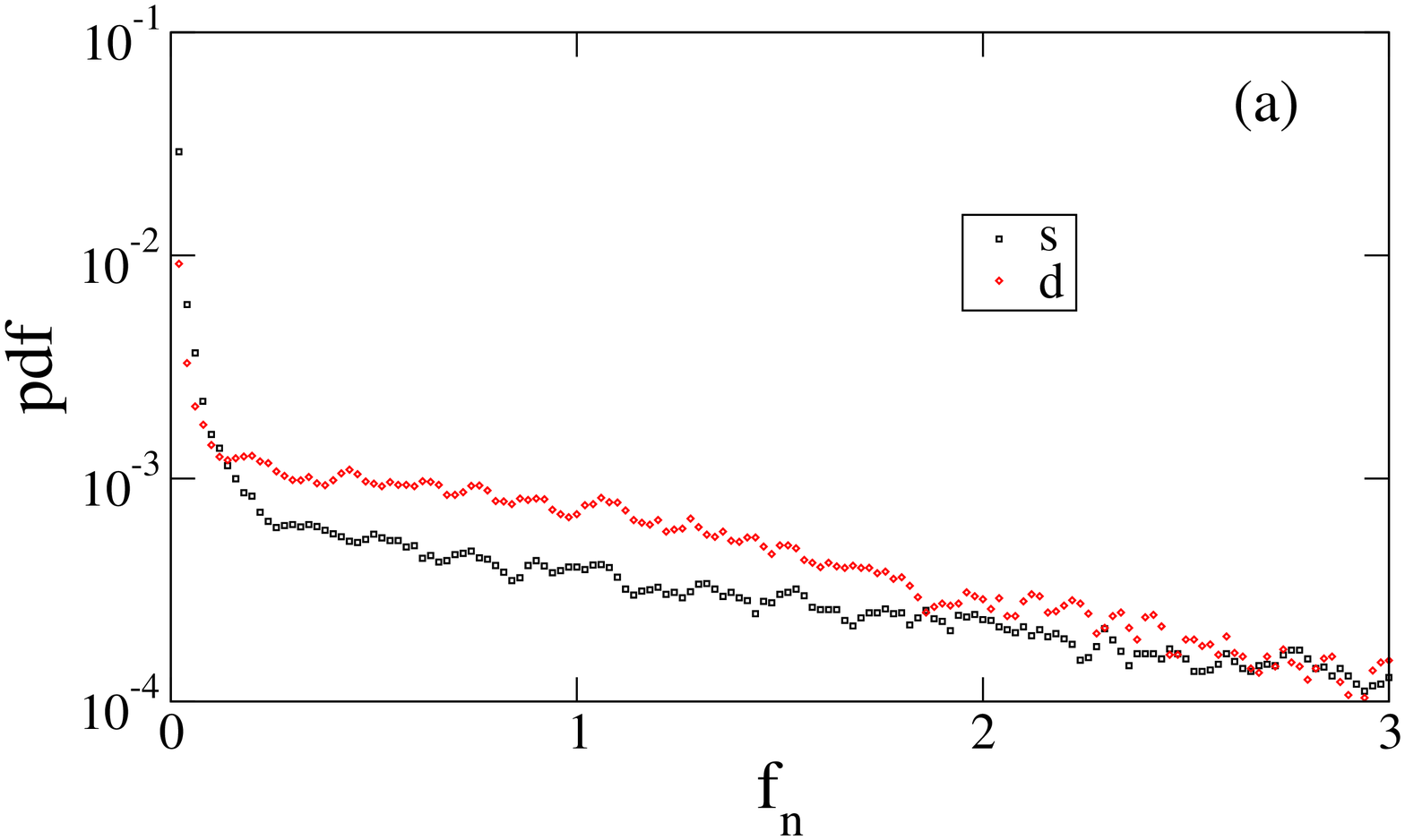}
\includegraphics[width=8cm]{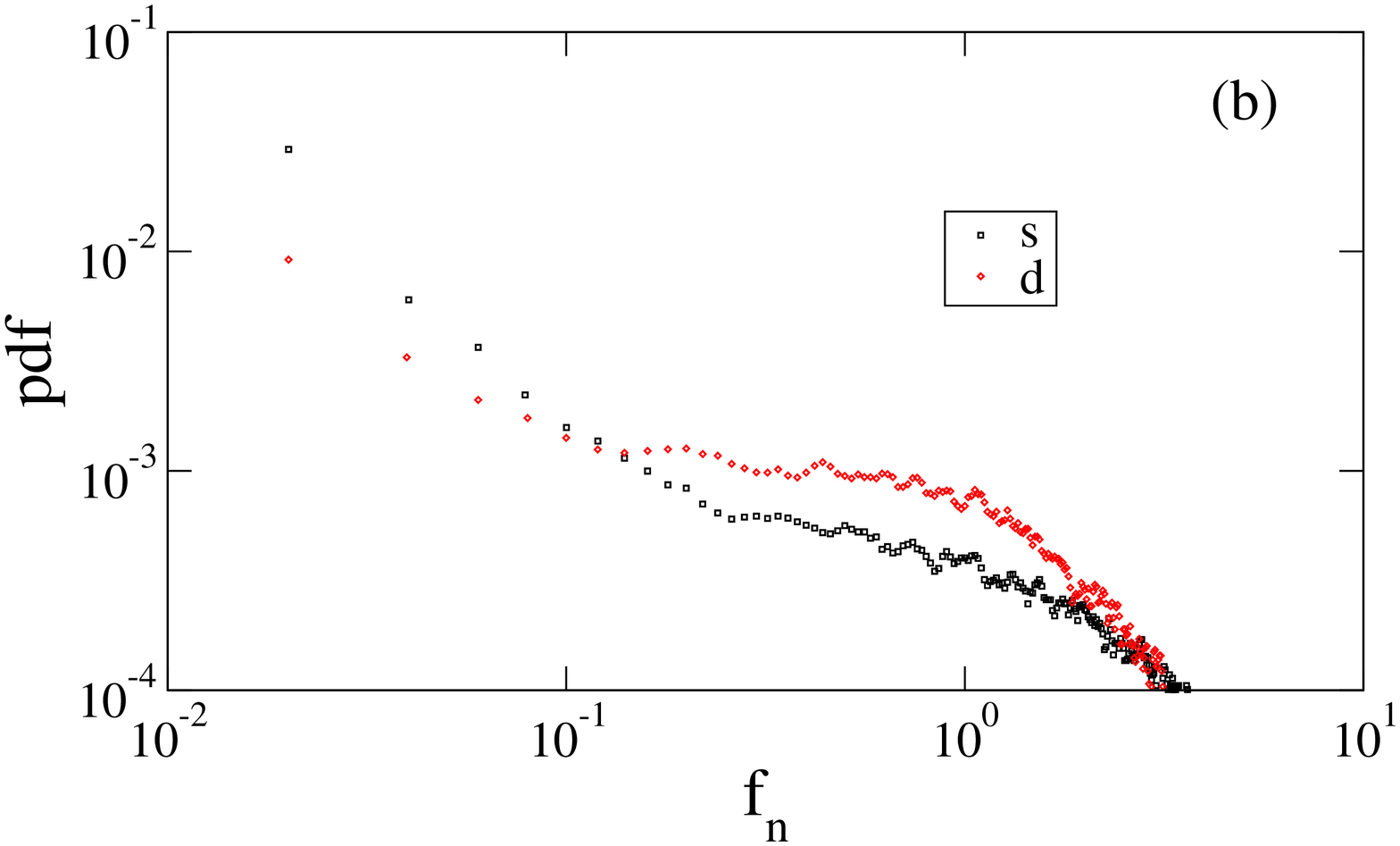}
\caption{Probability density function of normal forces for simple (s) 
and double (d) contacts in log-linear (a) and log-log scales (b).   
\label{fig23}}
\end{figure}

\begin{figure}
\includegraphics[width=8cm]{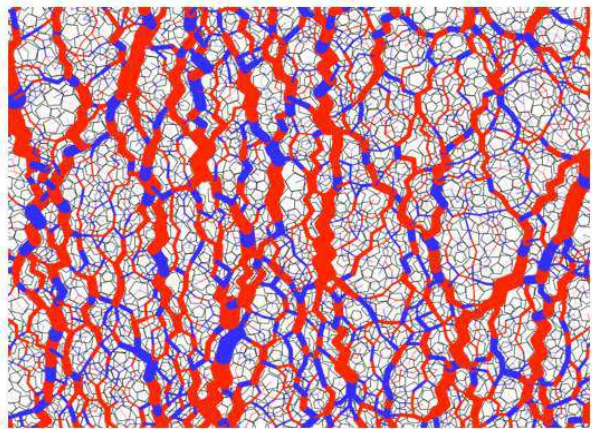}
\caption{(color online) Color map of the normal force network in the residual state with 
simple contacts (s) in blue and double contacts (d) in red. Line thickness is proportional to 
the normal force.    \label{fig24}}
\end{figure}

The normal force pdf's for simple and double contacts are shown in 
Fig. \ref{fig23}.  Both contact types are involved in weak and strong networks and  
the pdf's have the same functional form. 
But the contribution of simple contacts is more important in the range of 
very weak forces. Once again, as for anisotropy, the very weak contacts 
appear to be related to the particular geometry of the pentagons.   
At large strains, about $32\%$ of all contacts belong to the very weak force 
network with $25\%$ simple contacts against $7\%$  
double contacts. A snapshot of the normal force network is shown 
in Fig. \ref{fig24} where the line widths are proportional to 
the line width with different colors (or grey levels) for  simple and double contacts. 
The remarkable feature of this map is the network of  very strong zigzag force chains  
composed mostly of double contacts and occasionally mediated by simple contacts. 

\begin{figure}
\includegraphics[width=8cm]{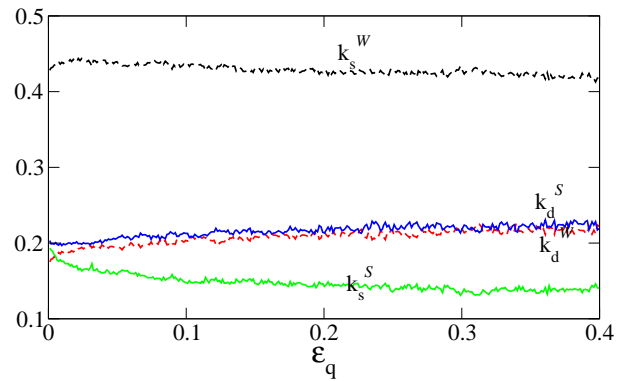}
\caption{ Proportions $k_s^S$ and $k_s^W$ of strong (S) and weak (W)  
simple (s) contacts, respectively, as well as the proportions $k_d^S$ and $k_d^W$ 
of strong and weak double (d) contacts  as a function of 
cumulative shear strain $\varepsilon_q$ in the pentagon packing.   \label{fig25}}
\end{figure}

The proportions $k_s^S$ and $k_s^W$ of strong (S) and weak (W)  
simple (s) contacts, respectively, as well as the proportions $k_d^S$ and $k_d^W$ 
of strong and weak double (d) contacts are plotted in Fig. \ref{fig25} as a function of 
$\varepsilon_q$. We see that in the strong network 
($f_n > \langle f \rangle$) the proportion $k_d^S$ of 
double contacts is nearly the same as the proportion $k_s^S$ of simple contacts in 
the initial (isotropic) state, but during shear $k_s^S$ declines down 
to $k_s^S \simeq 0.5 k_d^S$ in the residual state, in agreement with 
the impression left by  Fig. \ref{fig24}. We have an inverse situation 
for the weak network composed of two times more  simple 
contacts than double contacts, i.e.   $k_s^W \simeq 2 k_d^W$ 
in the residual state. It is also interesting to remark that the fraction of weak contacts, i.e. 
$k_s^W + k_d^W \simeq 0.58$ in the residual state is very close to that ($0.62$) in the case of the disk packing.       
 
\section{Conclusion}
\label{conclusion}

The objective of this paper was to isolate the effect of particle shape 
on force transmission in granular media by means a detailed comparison between 
two similar packings with different particle shapes: pentagons vs. disks. 
We observed enhanced shear strength and force inhomogeneity in 
the pentagon packing. But, unexpectedly, the pentagon packing was found to develop 
a lower structural (fabric) anisotropy compared to the disk packing under shear. 
This low fabric anisotropy, however,  does not  prevent the pentagon packing from  
building up a strong force anisotropy that underlies its enhanced shear strength 
compared to the disk packing. 

This finding is interesting as it shows unambiguously  that the force anisotropy in a 
granular material has two distinct sources: (1) Fabric anisotropy, with a 
maximum value depending on particle shapes; (2) Particle shapes. 
The first mechanism is crucial for the disk packing  so that the force anisotropy,  
and the shear stress as a result, vanishes in an isotropic disk packing (e.g. when 
the friction coefficient is set to zero). The second mechanism may be the 
predominant source of strength for ``facetted" particles that can give rise edge-to-edge (in 2D) or 
face-to-face (in 3D) contacts allowing for strong force localization along such contacts 
in the packing.  
Since the fabric anisotropy is low in a pentagon packing, the role of 
force anisotropy and thus the local equilibrium structures or arching are important 
with respect to its overall strength properties.  The pentagons analyzed in this work provide thus 
the first counter-example  of a system where the role of fabric anisotropy 
in shear strength is marginal.

Another shape-related effect was the observation of zig-zag force chains 
mostly composed of edge-to-edge contacts in steady shearing. The 
vertex-to-edge contacts belong thus mainly to the weak force network or a class 
of ``very weak'' forces that can be considered as a signature of 
enhanced arching or screening effect  of 
forces in the presence of  edge-to-edge chains. These ``very weak" forces can also 
be observed, though to a lesser extent, in a disk packing with high 
coefficients of friction \cite{Silbert2002} or on experimental pdf's of normal forces acting on 
the walls of a container \cite{Mueth1998a}. Let us recall that a ``very weak phase" was 
also evidenced by considering the correlation between friction mobilization 
and the anisotropy of granular texture in a disk packing at the stability limit \cite{Staron2005}.       

By focusing on pentagon packings, we were able to demonstrate the nontrivial 
phenomenology resulting from the specific shape of particles  as compared to 
a disk packing. Although general features of force 
transmission (pdf's, bimodal character, etc) seem to be 
robust, the details of force transmission (relative 
importance of force and fabric anisotropy, the role of 
edge-to-edge contacts, etc) seem to be strongly shape-dependent. Currently, we work  
to elucidate this issue for  regular polygons (hexagons and higher number of sides) as 
well as polyhedral particles in three dimensions.


We warmly thank Fr\'ederic Dubois for assistance with the LMGC90 platform 
used for the simulations. This work was funded 
by the French Railway Society, the SNCF, and the R\'egion Languedoc-Roussillon 
of France.  


\end{document}